# Broadband Frequency and Spatial On-Demand Tailoring of Topological Wave Propagation Harnessing Piezoelectric Metamaterials

**Patrick Dorin[1*], K.W. Wang[1]**

[1]Structural Dynamics and Controls Laboratory, Department of Mechanical Engineering, University of Michigan, Ann Arbor, Michigan, USA

**\* Correspondence:**
Patrick Dorin
pdorin@umich.edu

**Abstract**

Many engineering applications leverage metamaterials to achieve elastic wave control. To enhance the performance and expand the functionalities of elastic waveguides, the concepts of electronic transport in topological insulators have been applied to elastic metamaterials. Initial studies showed that topologically protected elastic wave transmission in mechanical metamaterials could be realized that is immune to backscattering and undesired localization in the presence of defects or disorder. Recent studies have developed tunable topological elastic metamaterials to maximize performance in the presence of varying external conditions, adapt to changing operating requirements, and enable new functionalities such as a programmable wave path. However, a challenge remains to achieve a tunable topological metamaterial that is comprehensively adaptable in both the frequency and spatial domains and is effective over a broad frequency bandwidth that includes a subwavelength regime. To advance the state of the art, this research presents a piezoelectric metamaterial with the capability to concurrently tailor the frequency, path, and mode shape of topological waves using resonant circuitry. In the research presented in this manuscript, the plane wave expansion method is used to detect a frequency tunable subwavelength Dirac point in the band structure of the periodic unit cell and discover an operating region over which topological wave propagation can exist. Dispersion analyses for a finite strip illuminate how circuit parameters can be utilized to adjust mode shapes corresponding to topological edge states. A further evaluation provides insight into how increased electromechanical coupling and lattice reconfiguration can be exploited to enhance the frequency range for topological wave propagation, increase achievable mode localization, and attain additional edge states. Topological guided wave propagation that is subwavelength in nature and adaptive in path, localization, and frequency is illustrated in numerical simulations of thin plate structures. Outcomes from the presented work indicate that the easily integrable and comprehensively tunable proposed metamaterial could be employed in applications requiring a multitude of functions over a broad frequency bandwidth.

## 1 Introduction

In recent years, it has been recognized that wave control utilizing elastic metamaterials can enhance performance and expand functionalities in many applications, such as energy harvesting, noise isolation, sensing, communications, filtering, and cloaking (Fang et al., 2018; Hussein et al., 2014; Wang et al., 2020). One objective of elastic metamaterials that has received significant attention is the confinement of elastic waves to a specified path or location through the formation of a waveguide. Conventional elastic waveguides obtain wave confinement by creating an inclusion where the wave can localize within a periodic lattice (Benchabane et al., 2005; Casadei et al., 2012; Kafesaki et al., 2000; Khelif et al., 2004; Liu et al., 2020; Oudich et al., 2010). These conventional waveguides can suffer from performance degradation when disorder (e.g., a sharp turn in the waveguide) or defects (e.g., a manufacturing imperfection) exist within the periodic lattice (Pal and Ruzzene, 2017; Sun and Wu, 2006). To avoid this performance degradation and expand the functionalities of elastic waveguides, the concepts that underly topologically protected conducting states in electronic materials (Hasan and Kane, 2010) have been employed in elastic metamaterials (Huber, 2016; Ma et al., 2019). Topological metamaterials are immune to various classes of disorder and defects that are oftentimes found in practical engineering applications, thus enabling lossless transmission along any desired wave path. To achieve topologically protected wave propagation that is localized to waveguides in two-dimensional systems, the elastic analogs of the quantum Hall effect (QHE), quantum spin Hall effect (QSHE), and quantum valley Hall effect (QVHE) have been extensively investigated.

Protected wave propagation that is confined to a waveguide is obtained through the activation of localized topological edge states that arise from the QHE, QSHE, and QVHE. Due to the active/moving components that are generally required to break time-reversal symmetry (TRS) for the QHE (Nash et al., 2015; von Klitzing, 1986; Wang et al., 2015a; Wang et al., 2015b) and the complex geometries necessary to achieve a double Dirac cone (a degeneracy where four cones meet) in the dispersion relation for the QSHE (Chaunsali et al., 2018; Kane and Mele, 2005; Miniaci et al., 2018; Mousavi et al., 2015; Süsstrunk and Huber, 2015; Wu and Hu, 2015), the relatively simpler QVHE has garnered significant attention. The QVHE requires the formation of a single Dirac cone (a degeneracy where two cones meet) in the dispersion relation of the unit cell. This single Dirac cone is opened with a lattice perturbation that breaks space inversion symmetry (SIS), which produces a bandgap that can support topological edge states due to valley-dependent topological properties (Nakada et al., 1996; Peres, 2010; Rycerz et al., 2007; Xiao et al., 2007). Topologically protected wave transmission according to the elastic analog of the QVHE has been obtained in reticular structures (Miniaci et al., 2019; Vila et al., 2017) and continuous thin plates with periodically placed masses (Chen et al., 2017; Yu et al., 2018; Zhu et al., 2018) or inclusions (Du et al., 2020). These structures emulate the elastic QVHE by exploiting a periodic impedance mismatch (Bragg scattering mechanism), and thus the resulting topological edge states exist at frequencies corresponding to wavelengths that are dependent on the lattice constant. The addition of local mass resonators (Lera et al., 2019; Pal and Ruzzene, 2017; Torrent et al., 2013; Wang et al., 2019; Zhang et al., 2020) and acoustic black holes (Ganti et al., 2020) to continuous plates has facilitated the achievement of topological edge states at frequencies that are determined by the characteristics of the local element (mass resonator or acoustic black hole). Topological wave propagation at low frequencies corresponding to wavelengths that are larger than the lattice constant has been demonstrated by carefully selecting the properties of these local elements (Chaunsali et al., 2018; Pal and Ruzzene, 2017; Wang et al., 2019; Zhang et al., 2020). Subwavelength topological metamaterials such as these could be highly valuable for engineering applications that require wave control at low frequencies in a size-constrained environment.

To extend beyond the functionalities available in fixed structures and enable adaptivity to varying operating requirements and external conditions, many recent investigations have focused on introducing tunability to topological metamaterials. The spatial path of topological wave propagation has been demonstrated to be adjustable by applying an external magnetic field (Q. Zhang et al., 2019), modifying mechanical boundary conditions (Darabi and Leamy, 2019; Tang et al., 2019), adding an elastic foundation (Al Ba'ba'a et al., 2020), connecting negative capacitance piezoelectric circuitry (Darabi et al., 2020; Riva et al., 2018), or switching stable states in bistable elements (Wu et al., 2018). The shape and localization of topological edge states have been tuned by exploiting an applied strain field (Liu and Semperlotti, 2018). Initial studies involving real-time frequency tuning of topological edge states have shown that an applied temperature (Liu et al., 2019), strain field (Nguyen et al., 2019), or electrical field (Zhou et al., 2020) can increase the frequency range over a limited region that is related to the Bragg mechanism or the magnitude of lattice perturbation.

The tunable topological metamaterials that have been investigated thus far have each concentrated on tailoring an individual characteristic of the topological wave to unlock novel functions and enhance performance in devices exhibiting elastic wave control. However, to fully realize the potential of topological metamaterials in wave control platforms and enable robust performance over a broad set of functionalities, an elastic metamaterial with multiple tunable topological characteristics must be developed. To date, concurrent tunability of the frequency and spatial characteristics of topological edge states has yet to be fully explored in a singular elastic metamaterial. Such a metamaterial would be of great benefit for devices such as wave filters, multiplexers, and energy harvesters that must route energy over a large frequency bandwidth. Besides primarily focusing on one adaptive characteristic, the currently established tunable topological metamaterials generally rely on the Bragg mechanism to generate a Dirac cone. Due to the interaction of platform-specific tuning parameters with the underlying physics of the Bragg mechanism, these metamaterials are either incapable of on-line frequency tuning (e.g., they are path tunable only) or can only practically do so over a limited range. Moreover, structures fabricated from these metamaterials would need to be large to control energy at low frequencies that correspond with fundamental system modes, which can be critical in structural applications. As a result, an easily integrable topological metamaterial that is capable of subwavelength elastic wave control and programmable over a broad frequency bandwidth has yet to be developed.

To advance the state of the art, the presented research proposes a piezoelectric topological metamaterial harnessing integrated resonant circuitry for comprehensively tunable subwavelength wave control. The goal of this investigation is to uncover insights into critical adaptive parameters and synthesize a framework for the attainment of adaptive topological wave propagation using resonant electromechanical metamaterials. In contrast to previous works, the rich tailorable characteristics of the proposed metamaterial enable the concurrent adaptation of the frequency range, path, and mode shape of topological edge states. This on-demand tunability of topological properties is achieved for the first time in a subwavelength (i.e., compact) and load-bearing thin plate structure.

To accomplish the research objective, this manuscript presents the evaluation of the dispersion relation for the unit cell from the plane wave expansion (PWE) method. This method is utilized to establish the working principle for the achievement of adaptive topological edge states from the QVHE. A parametric analysis is performed to identify and explore an achievable operating region for tunable topological wave propagation. Numerical computations of the dispersion relation for a finite strip of connected unit cells uncover how circuit parameters can enable the adjustment of topological edge states. Further analysis is conducted to investigate how topological edge state adaptivity can be augmented by the connection of negative capacitance to enhance electromechanical coupling and lattice reconfiguration via shorting circuits to obtain an additional Dirac cone. Finally, the activation of topological edge states for the achievement of guided topological wave propagation at lattice interfaces and boundaries is revealed by numerical simulations of a thin plate structure.

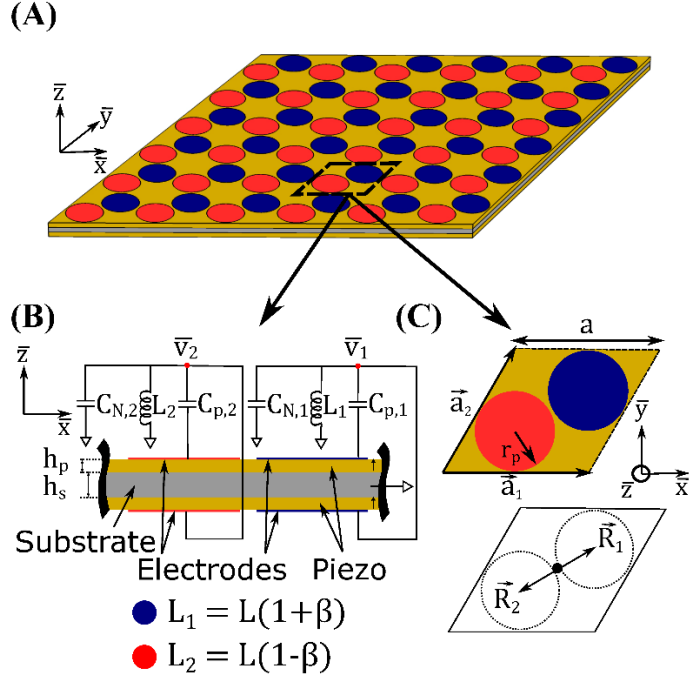

**Figure 1**: **(A)** Isometric view of piezoelectric metamaterial, with unit cell enclosed in dashed lines. **(B)** Cross-section and **(C)** top view of metamaterial unit cell. Blue indicates electrode geometry connected to circuit 1, red indicates electrode geometry connected to circuit 2.

## 2 Concept and Theoretical Model

### 2.1 System Description

As shown in **Figure 1**, the proposed piezoelectric metamaterial is comprised of a bimorph thin plate with substrate (gray) and piezoelectric (yellow) layers. The substrate layer has thickness $h_s$, density $\rho_s$, elastic modulus $E_s$, and Poisson's ratio $\nu_s$ while the piezoelectric layers have thickness $h_p$ and density $\rho_p$. The piezoelectric elements are connected to external circuitry through conductive circular electrodes (red and blue in **Figure 1**) with radius $r_p$ and a thickness that is assumed to be negligible (Zheng et al., 2019). The electrodes are arranged in a honeycomb lattice formation that contains the symmetries required to achieve the elastic analog of the QVHE. The metamaterial unit cell is defined by the basis vectors $\vec{a}_1 = a\hat{\imath}$ and $\vec{a}_2 = a\left(\cos\frac{\pi}{3}\hat{\imath} + \sin\frac{\pi}{3}\hat{\jmath}\right)$, where $a$ is the lattice constant, and the resulting unit cell area is $A_c = \frac{\sqrt{3}}{2}a^2$ (**Figure 1C**). This unit cell contains two electrode pairs that form capacitors (capacitor 1 and capacitor 2) that are centered at $\vec{R}_1 = \frac{a}{2\sqrt{3}}\left(\cos\frac{\pi}{6}\hat{\imath} + \sin\frac{\pi}{6}\hat{\jmath}\right)$ and $\vec{R}_2 = \frac{-a}{2\sqrt{3}}\left(\cos\frac{\pi}{6}\hat{\imath} + \sin\frac{\pi}{6}\hat{\jmath}\right)$. Each capacitor ($C_{p,1}$ and $C_{p,2}$) is connected to an inductor ($L_1$ and $L_2$) to form a resonant $LC$ circuit with a natural frequency of $\omega_{t,j} = \sqrt{\frac{1}{L_j C_{p,j}}}$ for the $j$th circuit. A negative capacitor with capacitance $C_{N,j}$ (i.e., for a capacitor with capacitance $C = -1$ F, $C_N = 1$ F) is connected in parallel to the $j$th resonant circuit. Ohmic losses are not considered in the derivation for the dispersion analysis, as all circuit elements are assumed to be ideal. In addition, damping is neglected, and perfect adhesion is assumed to exist between layers.

## 2.2   Governing Equations

In the derivation of the theoretical model for the metamaterial, small deflections and a thin structure (in the $\bar{z}$-direction) are assumed. The classical theory of thin plates (Graff, 1991) and the linear theory of piezoelectricity (Inman and Erturk, 2011; Tiersten, 2013) are applied. Assuming a transversely isotropic material and plane stress, the elastic, piezoelectric, and permittivity constants for the piezoelectric elements of the bimorph plate are obtained from the reduced (from the full three-dimensional form) piezoelectric constitutive relations for a thin plate (Inman and Erturk, 2011) as:

$$\begin{aligned}
\bar{c}_{11}^E &= \frac{s_{11}^E}{(s_{11}^E + s_{12}^E)(s_{11}^E - s_{12}^E)} \\
\bar{c}_{12}^E &= \frac{-s_{12}^E}{(s_{11}^E + s_{12}^E)(s_{11}^E - s_{12}^E)} \\
\bar{c}_{66}^E &= \frac{1}{s_{66}^E} = \frac{1}{2}(\bar{c}_{11}^E - \bar{c}_{12}^E) \\
\bar{e}_{31} &= \frac{d_{31}}{s_{11}^E + s_{12}^E} \quad \bar{\varepsilon}_{33}^S = \varepsilon_{33}^T - \frac{2d_{31}^2}{s_{11}^E + s_{12}^E}
\end{aligned} \tag{1}$$

where $\bar{c}_{11}^E$, $\bar{c}_{12}^E$, and $\bar{c}_{66}^E$ are the reduced elastic constants evaluated at a constant electric field, $\bar{e}_{31}$ is the reduced piezoelectric constant, $\bar{\varepsilon}_{33}^S$ is the reduced permittivity constant evaluated at a constant strain, $s_{11}^E$ and $s_{12}^E$ are elastic compliance constants, $d_{31}$ is a piezoelectric coefficient, and $\varepsilon_{33}^T$ is permittivity calculated under constant stress. A uniform electric field is assumed across each electrode. The governing equations for the flexural displacement and voltage response of the metamaterial are derived using the extended Hamilton's principle (Meirovitch, 1967), the integral form of Gauss's law (Inman and Erturk, 2011), and Kirchoff's laws, and are given as:

$$D_T \bar{\nabla}^4 \bar{w}(\bar{r},t) + m\frac{\partial^2 \bar{w}(\bar{r},t)}{\partial t^2} - \theta \sum_{j=1}^{N_e} \bar{\nabla}^2 \bar{v}_j(t)\bar{\chi}_j(\bar{r}) = 0 \tag{2}$$

$$L_j(C_{p,j} - C_{N,j})\frac{\partial^2 \bar{v}_j(t)}{\partial t^2} + \bar{v}_j(t) + \theta \iint_{\bar{D}_j} L_j \frac{\partial^2}{\partial t^2}\bar{\nabla}^2 \bar{w}(\bar{r},t)d^2\bar{r} = 0, \quad j = 1 \ldots N_e \text{ electrode pairs} \tag{3}$$

where $\bar{r} = (\bar{x}, \bar{y})$, $\bar{w}(\bar{r},t)$ is the flexural displacement of the plate, $m$ is the effective mass per unit area of the plate, $D_T$ is the effective flexural rigidity of the plate at short circuit, $\theta$ is an electromechanical coupling coefficient, $\bar{v}_j(t)$ is the output voltage across the $j$th electrode pair, $C_{p,j}$, $L_j$, and $C_{N,j}$ are the capacitance, connected inductance, and connected negative capacitance corresponding to the $j$th electrode pair, $N_e$ is the total number of electrode pairs, $\bar{\nabla}^2$ and $\bar{\nabla}^4$ are the Laplacian and biharmonic operators, respectively, and $t$ is time. Furthermore, $\bar{\chi}_j(\bar{r}) = \begin{cases} 1, & \bar{r} \in \bar{D}_j \\ 0, & otherwise \end{cases}$, where the value of the step-function $\bar{\chi}_j(\bar{r})$ depends on $\bar{D}_j$, which represents the subdomain in $\bar{x} - \bar{y}$ space containing the $j$th electrode pair. The definitions for $m$, $D_T$, $\theta$, and $C_{p,j}$ are given by:

$$m = \rho_s h_s + 2\rho_p h_p$$
$$D_T = D_s + D_p = \frac{E_s h_s^3}{12(1-\nu_s^2)} + \bar{c}_{11}^E \left(\frac{2h_p^3}{3} + h_p^2 h_s + \frac{h_p h_s^2}{2}\right) \tag{4}$$
$$\theta = \bar{e}_{31}(h_p + h_s)$$
$$C_{p,j} = \frac{2\bar{\varepsilon}_{33}^S}{h_p} A_{e,j} = \tilde{C}_p A_{e,j}$$

where $D_s$ and $D_p$ are the flexural rigidities of the substrate and the piezoelectric layers, respectively, $A_{e,j}$ is the area within the unit cell that contains the $j$th electrode pair, and $\tilde{C}_p$ is the capacitance per unit area. A harmonic response is assumed at frequency $\omega$, such that:

$$\bar{w}(\bar{r},t) = \bar{W}(\bar{r})e^{i\omega t} \quad \bar{v}_j(t) = \bar{V}_j e^{i\omega t} \tag{5}$$

To generalize the results, a nondimensionalization scheme is adopted and the resulting equations are:

$$\left(\nabla^4 - \frac{\omega^2 m a^4}{D_T}\right)w(r) - \sum_{j=1}^{N_e} \frac{\theta^2 a^2}{(C_{p,j} - C_{N,j})D_T} \nabla^2 v_j \chi_j(r) = 0 \tag{6}$$

$$\left(1 - L_j(C_{p,j} - C_{N,j})\omega^2\right)v_j - \omega^2 L_j(C_{p,j} - C_{N,j}) \iint_{D_j} \nabla^2 w(r)\, d^2 r = 0; \; j = 1 \ldots N_e \text{ electrode pairs} \tag{7}$$

where the non-dimensional flexural displacement, voltage, time, and length scales are defined as: $w = \frac{\bar{W}}{a}, v_j = \frac{1}{a}\frac{C_{p,j}-C_{N,j}}{\theta}\bar{V}_j, \tau = \sqrt{\frac{1}{L_j(C_{p,j}-C_{N,j})}}t, x = \frac{\bar{x}}{a}, y = \frac{\bar{y}}{a}$, and $z = \frac{\bar{z}}{a}$, respectively, $r = (x,y)$, $\nabla^2$ and $\nabla^4$ are the nondimensional Laplacian and biharmonic operators, respectively, and $\chi_j(r) = \begin{cases} 1, & r \in D_j \\ 0, & \text{otherwise} \end{cases}$, where $D_j$ is the subdomain in $x-y$ space containing the $j$th electrode pair.

## 2.3 Dispersion Relation

To attain the dispersion properties of the metamaterial, the unit cell is analyzed using the PWE method. For this study, the number of electrode pairs (capacitors) in the unit cell is set to $N_e = 2$, as is shown in **Figure 1**. The inductors $L_1$ and $L_2$ are defined as $L_1 = L(1 + \beta)$ and $L_2 = L(1 - \beta)$, where $\beta$ is a circuit inductance perturbation parameter. The PWE method is applied following the steps outlined in previous investigations utilizing mechanical resonators (Chaunsali et al., 2018; Pal and Ruzzene, 2017; Xiao et al., 2012), where the nondimensional flexural displacement $w(r)$ of the plate is expressed as a superposition of plane waves:

$$w(r) = \sum_G W(G)e^{-ia(k+G)\cdot r}, \; G = m\vec{b}_1 + n\vec{b}_2 \tag{8}$$
$$m, n \in [-M, M], \quad k = (k_x, k_y)$$

where $\vec{b}_1 = \frac{\pi}{a}\left(2\hat{\imath}, -\frac{2}{\sqrt{3}}\hat{\jmath}\right)$ and $\vec{b}_2 = \frac{\pi}{a}\left(0\hat{\imath}, \frac{4}{\sqrt{3}}\hat{\jmath}\right)$ are the reciprocal lattice basis vectors (a diagram of the reciprocal lattice is contained in the inset of **Figure 2**), $G$ is the reciprocal lattice vector, $m$ and $n$ are the plane wave indices, $k$ is the Bloch wavevector, and $M$ is an integer chosen such that $N^2 = $ $(2M + 1)^2$ is the number of plane waves that are included in the calculation. **Equation 8** is substituted into **Equation 6**, the result is multiplied by the complex conjugate $e^{ia(K+G')\cdot r}$, and integrated over the dimensionless unit cell $\iint_{A_{c-ND}} dr$, where $A_{c-ND} = \sqrt{3}/2$ represents the dimensionless area of the unit cell, defined as $A_{c-ND} = A_C/a^2$. Using the property of orthogonality:

$$\iint_{A_{c-ND}} e^{-ia(G-G')\cdot r} dr = \begin{cases} A_{c-ND} \text{ for } G = G' \\ 0 \text{ otherwise} \end{cases} \quad (9)$$

**Equation 10** is obtained. Similarly, by substituting **Equation 8** into **Equation 7**, **Equation 11** is obtained. The following equations define the dispersion relation of the metamaterial unit cell:

$$(a^4|k+G|^4 - \Omega^2)W(G) + \sum_{j=1}^{N_e} \frac{\vartheta}{1-\xi_j} \frac{a^2}{A_c} \frac{a^2}{A_{e,j}} a^2 |k+G|^2 \iint_{D_j} v_j e^{ia(k+G)\cdot r} d^2r = 0 \quad (10)$$

$$\left(\frac{\Omega_{t,j}^2}{(1-(-1)^j\beta)(1-\xi_j)} - \Omega^2\right) v_j + \Omega^2 a^2 \sum_G W(G)|k+G|^2 \iint_{D_j} e^{-ia(k+G)\cdot r} d^2r = 0 \quad (11)$$

$$j = 1 \ldots N_e \text{ electrode pairs}$$

where $\Omega$ is nondimensional frequency, $\vartheta$ is the nondimensional electromechanical coupling factor, $\Omega_{t,j}$ is the nondimensional circuit tuning frequency, and $\xi_j$ is the negative capacitance ratio for the $j$th circuit. The derived model is general and allows for unit cells with different electrode shapes and capacitance definitions. However, in this study, these features are selected to be uniform across both electrode pairs, as $C_{N,j} = C_N$, $C_{p,j} = C_p$, and $A_{e,j} = A_e$, such that $\Omega_{t,j} = \Omega_t$ and $\xi_j = \xi$. The equations for $\Omega$, $\vartheta$, $\Omega_t$, and $\xi$ are given as:

$$\begin{aligned} \Omega &= \omega a^2 \sqrt{\frac{m}{D_T}} \\ \vartheta &= \frac{\theta^2}{\tilde{C}_p D_T} \\ \Omega_t &= \sqrt{\frac{1}{LC_p}} a^2 \sqrt{\frac{m}{D_T}} \\ \xi &= \frac{C_N}{C_p} \end{aligned} \quad (12)$$

To evaluate the dispersion relation for the system, **Equations 10,11** are arranged in the form of the classical eigenvalue problem:

$$([\mathbf{K}] - \Omega^2[\mathbf{M}])[\mathbf{u}] = 0 \quad (13)$$

where the eigenvalues $\Omega^2$ and eigenvectors $[\mathbf{u}] = [\{W_{m,n}\} \quad v_1 \quad v_2]^T$ can be obtained by specifying the Bloch wavevector $k$. The matrices $[\mathbf{K}]$ and $[\mathbf{M}]$ are explicitly defined in Section 1 of the Supplementary Material.

## 2.4 Negative Capacitance Circuitry

Per **Equation 12,** $\Omega_t$ and $\xi$ are parameters that are directly controllable with circuit elements (e.g., synthetic inductor or negative capacitor), while the electromechanical coupling factor $\vartheta$ is dependent on material and geometric properties, and thus cannot be altered after metamaterial fabrication. While $\vartheta$ cannot be controlled directly by external circuitry, the addition of a negative capacitor in parallel to resonant circuitry has been shown to effectively (i.e., "synthetically") enhance the electromechanical coupling of a piezoelectric system. Previous studies have exploited an enhanced electromechanical coupling through negative capacitance circuitry to achieve vibration attenuation over a broad frequency bandwidth (Berardengo et al., 2016; Hagood and von Flotow, 1991; Sugino et al., 2017; Tang and Wang, 2001; Wang and Tang, 2008). In this investigation, negative capacitance circuitry is also used to tailor the electromechanical coupling of the system, as can be seen by defining an effective electromechanical coupling factor $\vartheta_{eff}$ as:

$$\vartheta_{eff} = \frac{\vartheta}{1-\xi} \qquad (14)$$

This term, which is present in **Equation 10**, measures the level of effective electromechanical coupling when accounting for the connected negative capacitor. Thus, by careful selection of the negative capacitance ratio (e.g., $\xi \approx 1$), the effective electromechanical coupling $\vartheta_{eff}$ can be significantly enhanced.

Since $C_N$ is an active component, a stability analysis is necessary. The stability requirement is obtained from the dispersion relation for the metamaterial by evaluating the eigenvalues of **Equation 13**, per the technique outlined in (Hu et al., 2020). For positive eigenvalues ($\Omega^2 > 0$), the oscillation frequencies ($\pm\omega$) are purely real, and there is a bounded oscillatory response (see **Equation 5**). On the other hand, for negative eigenvalues ($\Omega^2 < 0$), the result is frequencies with a nonzero imaginary part ($\pm i\omega$) and an unbounded response (**Equation 5**). For the proposed metamaterial, when $\xi < 1$, all eigenvalues are positive, and the system response is bounded. In contrast, when $\xi > 1$, negative eigenvalues appear, and the system response is unbounded. Therefore, to maintain stability:

$$C_N < C_p \leftrightarrow \xi < 1 \qquad (15)$$

There is a physical explanation for the derived stability criterion. The parallel-connected negative capacitor reduces the effect of the inherent capacitance of an electrode pair, causing a reduction in the total capacitance present in the corresponding circuit ($C_T = C_p - C_N$, see **Equation 3**). $C_T$ must remain positive to maintain stability, since negative total capacitance in a circuit is analogous to negative compliance (stiffness) in a mechanical system, and no additional balancing terms exist in the circuit governing equation (**Equation 3**).

In addition to changing the effective electromechanical coupling, the inclusion of negative capacitance influences the effective nondimensional tuning frequency of the resonant circuits. Through observation of **Equation 11**, this influence can be measured as:

$$\Omega_{t-eff}^2 = \frac{\Omega_t^2}{1-\xi} \qquad (16)$$

where $\Omega_{t-eff}$ is the effective nondimensional tuning frequency. Therefore, as the negative capacitance ratio $\xi$ is increased towards unity and effective coupling is enhanced, the effective tuning frequency is shifted to a higher value. The result indicates that, if $\xi$ is specified to be close to unity, a large inductance would be required to achieve a low-frequency value for $\Omega_{t-eff}$ because $\Omega_t$ would need to be set to a very low value. If the inductance required to achieve a desired low tuning frequency is too large to be achieved with a standard inductor, a synthetic inductor created from active circuit components (Kumar and Shukla, 1989) or a specially designed large passive inductor (Lossouarn et al., 2017) could be utilized. If these alternative inductor solutions are deemed impractical due to power or size requirements, an alternative for enhancing the effective electromechanical coupling is to connect the negative capacitor to the resonant circuit in a series configuration (Tang and Wang, 2001; Wang and Tang, 2008), instead of parallel (see Supplementary Material Section 2).

## 3 Working Principle – Obtainment of Tunable Topological Wave Propagation

In this section, the working principle is outlined for the attainment of subwavelength topological wave propagation using the proposed metamaterial. A unit cell dispersion analysis is conducted to identify lattices with nontrivial topological characteristics and define metrics that are indicators of waveguide performance. A finite strip analysis is undertaken to investigate how localized topological edge states can be obtained through the connection of topologically distinct lattices. Numerical simulations are performed to illustrate how these edge states can be exploited to achieve guided topological wave propagation that is tunable in both the frequency and spatial domains.

### 3.1 Unit Cell Dispersion Analysis

The band structure (i.e., the dispersion diagram) of the proposed metamaterial is obtained through a unit cell dispersion analysis. For this study, PZT-5H (a commonly utilized piezoceramic) is selected as the material for the piezoelectric layers and aluminum is selected as the material for the substrate layer. The material properties associated with PZT-5H and aluminum that are used in dispersion calculations are listed in **Table 1**. The geometric dimensions specified for the analysis are also included in **Table 1**. The layer thicknesses $h_s$ and $h_p$ are selected such that the nondimensional electromechanical coupling factor is maximized for the selected materials ($\vartheta = 0.42$, see **Equation 12**). The negative capacitance ratio is set to $\xi = 0.79$, such that the effective electromechanical coupling factor $\vartheta_{eff}$ is equal to 2. The tuning frequency of the circuit ($\Omega_t$) is selected as 7.3, resulting in an effective tuning frequency ($\Omega_{t-eff}$) of 16.0. The dispersion analysis of the unit cell is performed by solving the eigenvalue problem derived from the PWE method (**Equation 13**). $M = 3$ is chosen, such that $N^2 = (2M + 1)^2 = 49$ plane waves are considered in the calculation. The eigenvalues $\Omega^2$ are obtained by specifying the Bloch wavevector $k$ to follow the edges of the irreducible Brillouin zone (IBZ), which is shown as the blue triangle in the inset of **Figure 2**. The band structure for the unit cell with both inductance parameters set as identical values (i.e., $\beta = 0$) is displayed as the solid red lines in **Figure 2.** The band structure is also generated using the commercial finite element (FE) software COMSOL Multiphysics to validate the results derived from the PWE method (see Section 3 of Supplementary Material for more information on the FE model). Comparison of the results (in **Figure 2**) generated from the PWE method (solid red lines) and FE simulations (open black circles) indicates a close match, despite the solution from the PWE method requiring multiple orders of magnitude fewer degrees of freedom (13348 for FE vs. 51 for PWE).

Due to the presence of $C_3$ lattice symmetry, SIS, and TRS (Ma et al., 2019), a Dirac point (which is the vertex of a Dirac cone in $k_x$- $k_y$ space) is formed between the first and second bands at the $K$ point in reciprocal space (see black box in **Figure 2**). The Dirac point exists at the Dirac frequency $\Omega_{Dirac} = 8.9$. Per the QVHE, topological wave propagation can be obtained at frequencies near this Dirac point. To determine whether subwavelength topological wave propagation could be obtained at this frequency, the band structure for a bare bimorph plate (defined to be the bimorph plate consisting of the substrate layer, piezoelectric layers, and electrodes, with all of the connected circuits shorted) is calculated using the PWE method (dashed gray curves in **Figure 2**). At the Dirac frequency, the wavelength of propagating waves in the bare plate is $\lambda_b = 97$ mm (marked by a black star in **Figure** 2), which is 2.4 times larger than the lattice constant ($a = 40$ mm). Thus, by connecting resonant circuitry in the proposed metamaterial, the Dirac point is attainable in a subwavelength frequency regime. This subwavelength characteristic could be leveraged in applications that require low frequency (corresponding to large wavelengths) topological wave control in a small package.

**Table 1:** Definition of geometric and material properties.

| Parameter | Value |
|---|---|
| *Geometric Dimensions* | |
| $a$ | 40 mm |
| $h_s$ | 1 mm |
| $h_p$ | 1 mm |
| $r_p$ | 10.6 mm |
| *Substrate Layer Material Properties* | |
| $\rho_s$ | 2700 kg/m$^3$ |
| $E_s$ | 70 GPa |
| $\nu_s$ | 0.3 |
| *Piezoelectric Layer Material Properties* | |
| $\rho_p$ | 7500 kg/m$^3$ |
| $\bar{c}_{11}^E$ | 66.2 GPa |
| $\bar{e}_{31}$ | -23.4 C/m$^2$ |
| $\bar{\varepsilon}_{33}^S$ | 17.3 nF/m |

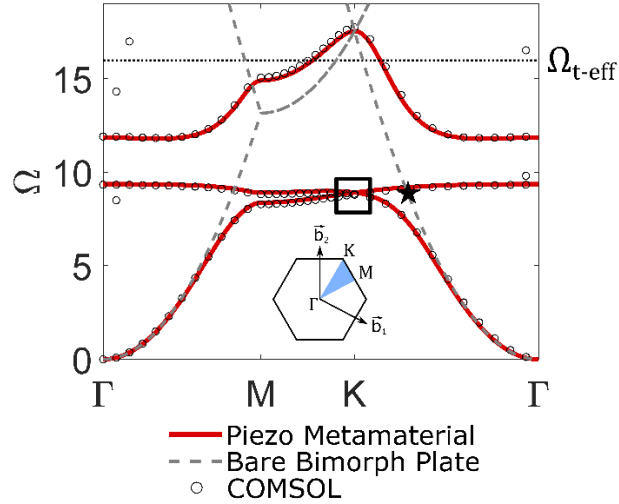

**Figure 2:** Dispersion diagram for unit cell with $\beta$ = 0 and $\Omega_{t-eff}$ = 16.0. The red lines represent results from PWE method, and the open black circles represent results from FE simulations. Band structure for bare bimorph plate is shown as gray dashed lines. Dirac point is enclosed in the black box. Inset contains schematic of reciprocal lattice and IBZ (blue triangle).

To obtain topological edge states per the QVHE, different inductance values are selected for each of the two circuits in the unit cell ($\beta \neq 0$), which breaks SIS. The band structure for $\beta = \pm 0.04$ is shown as the dotted lines in **Figure 3A** (the band structure for $\beta = 0$ is also included as solid lines for reference). Due to the broken SIS when $\beta \neq 0$, a full bandgap (i.e., topological bandgap) is opened from the Dirac point, which is indicated by the shaded region ($\Omega_{bandgap}$). To provide a measure of bandgap size that is not skewed toward higher frequencies, a relative bandgap ($\Omega_{bandgap-relative}$) is defined as:

$$\Omega_{bandgap-relative} = \frac{\Omega_{2-min} - \Omega_{1-max}}{\frac{\Omega_{2-min} + \Omega_{1-max}}{2}} = \frac{2|\Omega_{bandgap}|}{\Omega_{2-min} + \Omega_{1-max}} \quad (17)$$

where $|\Omega_{bandgap}|$ is the magnitude of $\Omega_{bandgap}$ (shown in **Figure 3A**), and $\Omega_{p-min}$ and $\Omega_{p-max}$ represent the minima and maxima of the $p$th band, respectively. A lattice with the inductance perturbation parameter specified as $\beta > 0$ is defined as a Type A lattice, whereas $\beta < 0$ for a Type B lattice (see schematics in **Figure 3B**). While the dispersion diagrams for Type A ($\beta = 0.04$) and Type B ($\beta = -0.04$) lattices are identical (**Figure 3A**), a band inversion exists between the two lattice types. The mode shapes for the first two bands evaluated at the $K$ point are shown in **Figure 3C** for Type A and Type B lattices. These mode shapes illustrate the band inversion, as the eigenvectors $u_p(k)$ for the $p$th band (where $p = 1$ is marked by a triangle and $p = 2$ is marked by a square) are interchanged for Type A and Type B lattices. This band inversion contributes to different topological characteristics for each lattice type. These topological characteristics are quantified by evaluating the topological invariant for the QVHE, the valley Chern number $C_{v-p}$, which is defined as (Berry, 1984; Xiao et al., 2007; Zhu et al., 2018):

$$C_{v-p} = \frac{1}{2\pi} \iint_v B_p(k) d^2k$$

$$B_p(k) = -\nabla_k \times \langle u_p(k)|i\nabla_k[\mathbf{M}]|u_p(k)\rangle$$

(18)

where $C_{v-p}$ is the topological charge (called the valley Chern number) for the $p$th band, which is calculated by integration of the Berry curvature ($B_p(k)$) near the $K$ point in reciprocal space. The theoretical values for $C_{v-1}^{Type\,A}$, $C_{v-2}^{Type\,A}$, $C_{v-1}^{Type\,B}$, and $C_{v-2}^{Type\,B}$ are -0.5, 0.5, 0.5, and -0.5, respectively (Yao et al., 2009; Zhu et al., 2018). The opposite signs of the $C_{v-p}$ values for Type A and Type B lattices indicate that they are topologically distinct. A method that is commonly utilized to obtain topological edge states is the connection of two topologically distinct lattices (e.g., Type A and Type B lattices) at an interface. The result is a topological transition and $N_{interface-states}$ (where $N_{interface-states} = |C_{v-p}^{Type\,A} - C_{v-p}^{Type\,B}|$) protected edge states with displacement localized at the interface, otherwise referred to as topological interface states (Ma et al., 2019; Yao et al., 2009). Therefore, a topological interface state can be obtained by connecting the Type A and Type B lattices outlined in this analysis ($N_{interface-states} = 1$).

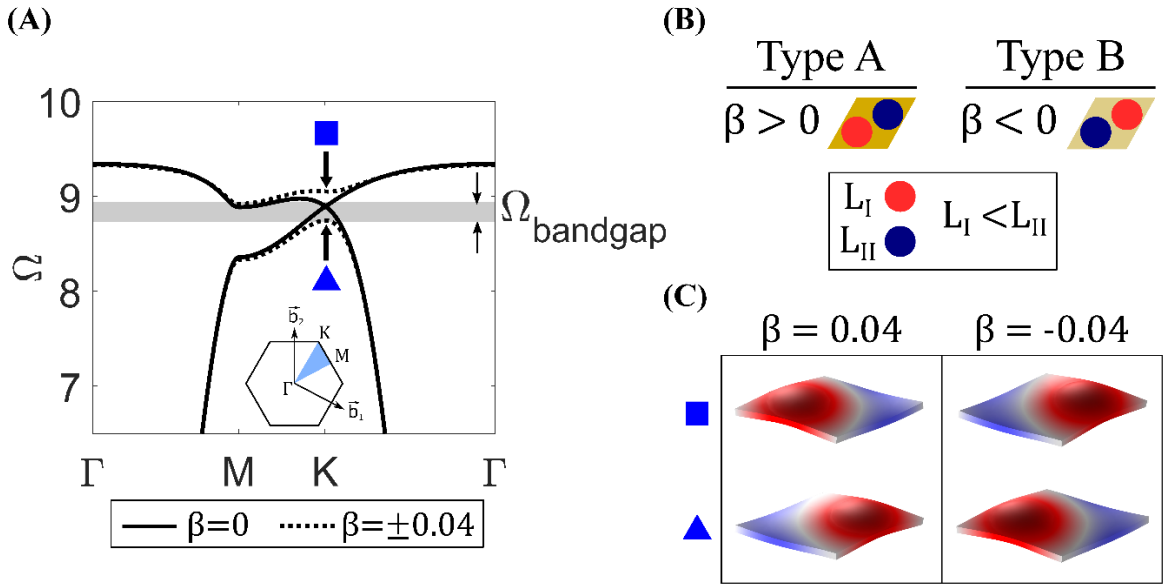

**Figure 3:** (**A**) Dispersion curves for $\beta = 0$ (solid lines) and $\beta = \pm 0.04$ (dotted lines) with $\Omega_{t-eff} = 16.0$. The full bandgap $\Omega_{bandgap}$ opened from the Dirac point when $\beta = \pm 0.04$ is indicated by the shaded region. (**B**) Schematic of unit cells for Type A ($\beta > 0$) and Type B ($\beta < 0$) lattices. The smaller inductance ($L_I$) is connected to the red electrode and the larger inductance ($L_{II}$) is connected to the blue electrode. (**C**) Mode shapes evaluated at the $K$ point for the first band (▲) and second band (■), revealing a band inversion between Type A ($\beta = 0.04$) and Type B ($\beta = -0.04$) unit cells.

The topological interface state must be excited in a way that does not activate the bulk modes of a given structure to obtain localized topological wave propagation. To achieve this goal, the interface state is excited at a frequency within the common topological bandgap that is found in the dispersion diagrams of the Type A and Type B lattices ($\Omega_{bandgap}$, **Figure 3A**). Therefore, the potential operating bandwidth of the topological interface state for a fixed selection of system parameters spans the topological bandgap ($\Omega_{bandgap}$). A nontrivial topological bandgap is required for an easily activated interface state to exist, and the larger the bandgap is, the greater the potential operating bandwidth of the topological waveguide. In addition to a nontrivial operating bandwidth, topological protection from defects and disorder is a required characteristic of the topological interface state. According to the elastic analog of the QVHE, the robustness to defects and disorder is related to the localization of the Berry curvature $B_p(k)$ at the $K$, $K'$ points in reciprocal space (Du et al., 2020; Qian et al., 2018). The magnitude of the valley Chern number $|C_{v-p}|$ provides a measure of this localization and thus describes the level of topological protection inherent to the interface state. The closer the value of $|C_{v-p}|$ is to the theoretical value of 0.5, the greater the amount of topological protection. The value of 0.5 is an idealized value for $|C_{v-p}|$, as the lattice perturbation that breaks SIS and opens the topological bandgap reduces its magnitude (Nguyen et al., 2019; Qian et al., 2018). Previous investigations into the elastic analog of the QVHE have shown that $|C_{v-p}| \geq 0.25$ can provide a sufficient amount of topological protection from disorder and defects in mechanical lattices (Nguyen et al., 2019; Zhu et al., 2018).

Based on this discussion, the performance criteria defined for this investigation that are obtainable from the unit cell dispersion analysis are (1) $\Omega_{bandgap-relative} > 0$, such that a nontrivial potential operating bandwidth exists, and (2) $|C_{v-p}| \geq 0.3$ for $p = 1,2$, such that there is a minimum acceptable level of topological protection. The values of $\Omega_{bandgap-relative}$ and $|C_{v-p}|$ are 0.02 and 0.3, respectively, for the parameters specified in this analysis (where $\Omega_{t-eff} = 16.0$ and $\beta = \pm 0.04$).

These results indicate that a topologically protected edge state would emerge in a structure containing an interface between the Type A ($\beta = 0.04$) and Type B ($\beta = -0.04$) lattices.

## 3.2 Topological Interface States

A dispersion analysis is conducted for a finite strip of unit cells containing an interface between Type A and Type B lattices to demonstrate the emergence of topologically protected interface states. For this finite strip analysis, all parameters are defined identically to the unit cell analysis discussed in the previous section ($\beta = \pm 0.04$, $\Omega_{t-eff} = 16.0$, $\vartheta_{eff} = 2$, and all parameters from **Table 1**). The finite strip is comprised of 18 unit cells, with nine Type A unit cells connected to nine Type B unit cells at an interface (see **Figure 4** for a schematic). A periodic boundary condition is applied in the $k_{//}$ direction, while the remaining boundaries at the ends of the finite strip are fixed. An interface composed of adjacent smaller inductance values ($L_I$, marked in red on the schematic) is designated as a Type I interface (**Figure 4A**), while an interface composed of adjacent larger inductance ($L_{II}$, marked in blue on the schematic) values is designated as a Type II interface (**Figure 4B**). The dispersion diagrams for Type I (**Figure 4A**) and Type II (**Figure 4B**) interfaces are generated using COMSOL Multiphysics. For each band, a localization parameter $\lambda$ is defined to measure the amount of flexural displacement that is localized at the interface as:

$$\lambda = \frac{\iiint_{V_{interface}} |w|^2 dV}{\iiint_{V_S} |w|^2 dV} \tag{19}$$

where $V_{interface}$ is the total volume of the two adjacent unit cells at the interface (enclosed in the dashed black boxes in **Figure 4**) and $V_S$ is the volume of the entire finite strip. This localization parameter is calculated for each band and is represented in the dispersion diagrams as a colormap. For a mode shape with flexural displacement that is highly localized to the interface (i.e., an interface state), the band is a dark red shade. Lighter shaded bands indicate bulk modes ($\lambda \ll 1$). The rectangular gray shaded region in both **Figure 4A** and **Figure 4B** separates an acoustic (low frequency) set of bulk modes from an optical (higher frequency) set of bulk modes and closely aligns with the topological bandgap ($\Omega_{bandgap}$) calculated in the unit cell dispersion analysis. For the Type I interface (**Figure 4A**), an interface state with highly localized displacement emerges from the optical set of bulk bands and crosses the topological bandgap to the acoustic set of bulk bands. A mode shape corresponding to this interface state at $\Omega = 8.5$ is shown in **Figure 4A,** where symmetric localized displacement exists at the interface. This mode shape contains the maximum localization of flexural displacement at the interface (as measured by $\lambda_m = 0.92$) when compared to all other interface state modes that reside within the topological bandgap. The mathematical expression for $\lambda_m$ is given as:

$$\lambda_m = \max_{\Omega \in \Omega_{bandgap}} \lambda(\Omega) \tag{20}$$

An additional localized state exists near $\Omega = 9.4$. However, this interface state is difficult to utilize without activating the bulk modes since it is not in the topological bandgap. In contrast to the Type I interface, for a Type II interface, the primary interface state emerges from the acoustic bulk bands and crosses the topological bandgap to approach the optical bulk bands (**Figure 4B**). A mode shape corresponding to this interface state at $\Omega = 8.7$ is shown in **Figure 4B**, where the flexural displacement field is now antisymmetric about the interface, and the resulting maximum displacement localization is somewhat less ($\lambda_m = 0.72$ compared to $\lambda_m = 0.92$ for the Type I interface state).

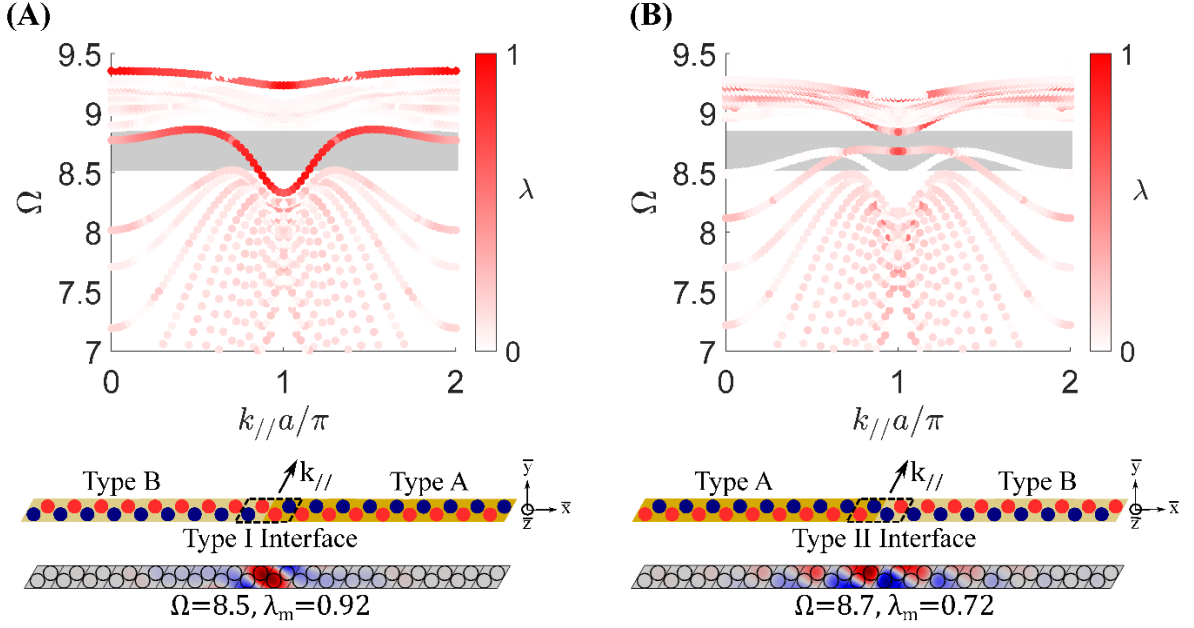

**Figure 4:** **(A)** Band structure for a finite strip ($|\beta| = 0.04$, $\Omega_{t-eff} = 16.0$, $\vartheta_{eff} = 2$, and all parameters from Table 1) with a Type I interface. The colormap indicates the localization of the flexural displacement at the interface through the localization parameter $\lambda$, with darker red shading indicating localized interface states ($\lambda \approx 1$). The rectangular gray shaded region represents a frequency range where no bulk modes exist and corresponds to the topological bandgap. The diagram of the finite strip is shown below the band structure, with the interface used for $\lambda$ calculations enclosed in a dashed black box. A symmetric mode shape that is calculated from the interface state at $\Omega = 8.5$ with a localized displacement ($\lambda_m = 0.92$) is also shown. **(B)** Band structure and schematic for a finite strip with a Type II interface. At the bottom, an antisymmetric mode shape that is calculated from the interface state at $\Omega = 8.7$ (with $\lambda_m = 0.72$).

Two observations are gained from the finite strip analysis. First, the proposed metamaterial enables the obtainment of symmetric and antisymmetric topological interface states, which aligns with previous investigations into the QVHE (Zhu et al., 2018). Due to the tunability of the circuit parameters in the proposed metamaterial, switching between these interface state types (symmetric and antisymmetric) could easily be achieved in practice. Second, the finite strip dispersion analysis supports the performance requirement for a nontrivial topological bandgap ($\Omega_{bandgap-relative} > 0$) derived from the unit cell dispersion analysis. In the finite strip analysis, the potential frequency bandwidth for both interface state types aligns with the topological bandgap, as unwanted hybridization of the interface state with bulk states may occur for frequencies outside of the targeted bandgap range (the rectangular gray shaded regions in **Figure 4**).

### 3.3 Path Tunable Topological Wave Propagation

The ability to achieve topological wave propagation from the interface states is investigated with FE simulations of a plate constructed from the proposed metamaterial. The proposed metamaterial enables concurrent tunability of the topological wave path, mode shape, and frequency, which advances upon previously developed platforms that are generally narrowband in nature and only focus on one tailorable characteristic. Each of the proposed metamaterial's tailorable properties are investigated separately in this manuscript to simplify the analysis. The path tunability of the topologically protected waveguide is examined in this section, while comprehensive analyses on frequency and mode shape tailoring are contained in section **Parametric Study – Frequency and Mode Shape Tunability.**

The plate used for FE simulations consists of a 16 by 20 lattice of unit cells, with low reflection conditions applied along all outer boundaries to suppress reflections (see left column of **Figure 5** for plate schematics). The lattice contains an interface between Type A and Type B unit cells, which is enclosed by the black lines in **Figures 5A,B,C,D**. A resistance ($R$) of 10 Ω is applied to each individual circuit to account for minor circuit losses that may arise in practical implementation. The plate is excited harmonically at a frequency that is within the topological bandgap ($\Omega_e = 8.7$), thus corresponding to a subwavelength regime (since the Dirac point at $\Omega_{\text{Dirac}} = 8.9$ contains the subwavelength characteristic). The excitation is applied as an out-of-plane point excitation at the location indicated by the arrow. FE simulations using COMSOL Multiphysics are conducted to obtain the steady-state displacement field. **Figure 5A** shows a displacement field with flexural response that is guided along a Type I interface, which supports a symmetric interface state. This specific case shows a straight line of wave propagation that is localized at the interface and is guided to a receiver indicated in **Figure 5A** as "R1". In **Figures 5B,C**, the interface is changed using circuit parameters such that flexural response under the same excitation is guided to receivers "R2" or "R3". In **Figure 5B,** guided transmission is demonstrated along a Type I interface with a sharp ($120°$) corner, indicating that the nontrivial topological property of the interface state provides protection from disorder. In **Figure 5C,** a Type I interface and a Type II interface are concatenated to achieve topological wave propagation around a shallow ($60°$) corner. It can be seen that the flexural displacement is symmetric along the Type I interface segment and antisymmetric along the Type II segment, revealing that the two edge state types are compatible and can be used in series to achieve a variety of interface paths. The results in **Figures 5A,B,C** illustrate the path tunability of the proposed topological metamaterial. As shown in the presented example, an input can be guided to three different receivers by using circuit elements (e.g., a synthetic inductor that can tune $\beta$) to vary the interface properties and location. For each of these cases, the wave amplitude (as indicated by the color intensity) is nearly identical from the input location to the designated output receiver, while the energy present at the other output receivers is negligible (e.g., in **Figure 5A**, there is a trivial amount of flexural displacement present near R2 and R3). This realization of nearly lossless transmission to the desired location(s) is obtainable due to the robustness of the topological interface state and the presence of the topological bandgap, which minimizes energy leakage into the bulk. Finally, to demonstrate the robustness of the topological interface state in the presence of both disorder and defects, a lattice is constructed with two sharp corners and a defect (one unit cell with both circuits shorted) in the interface path (**Figure 5D**). The steady-state response of the plate reveals flexural displacement that is guided along the Z-shaped interface path without undesirable localization or amplitude reduction at defects or sharp corners. This result indicates that guided transmission through more complicated wave paths (e.g., having multiple corners) can be obtainable from the proposed metamaterial, even in the presence of imperfections (e.g., malfunctioning/shorted circuits) that are commonly found in practical applications.

## 4    Parametric Study – Frequency and Mode Shape Tunability

In this section, a parametric study is undertaken to extensively explore the adaptive characteristics of the proposed metamaterial and develop a detailed understanding of system parameters that govern performance. A framework is developed to assess the frequency and interface mode shape tunability of the metamaterial and uncover insights into the impact of electromechanical coupling on topological elastic wave control.

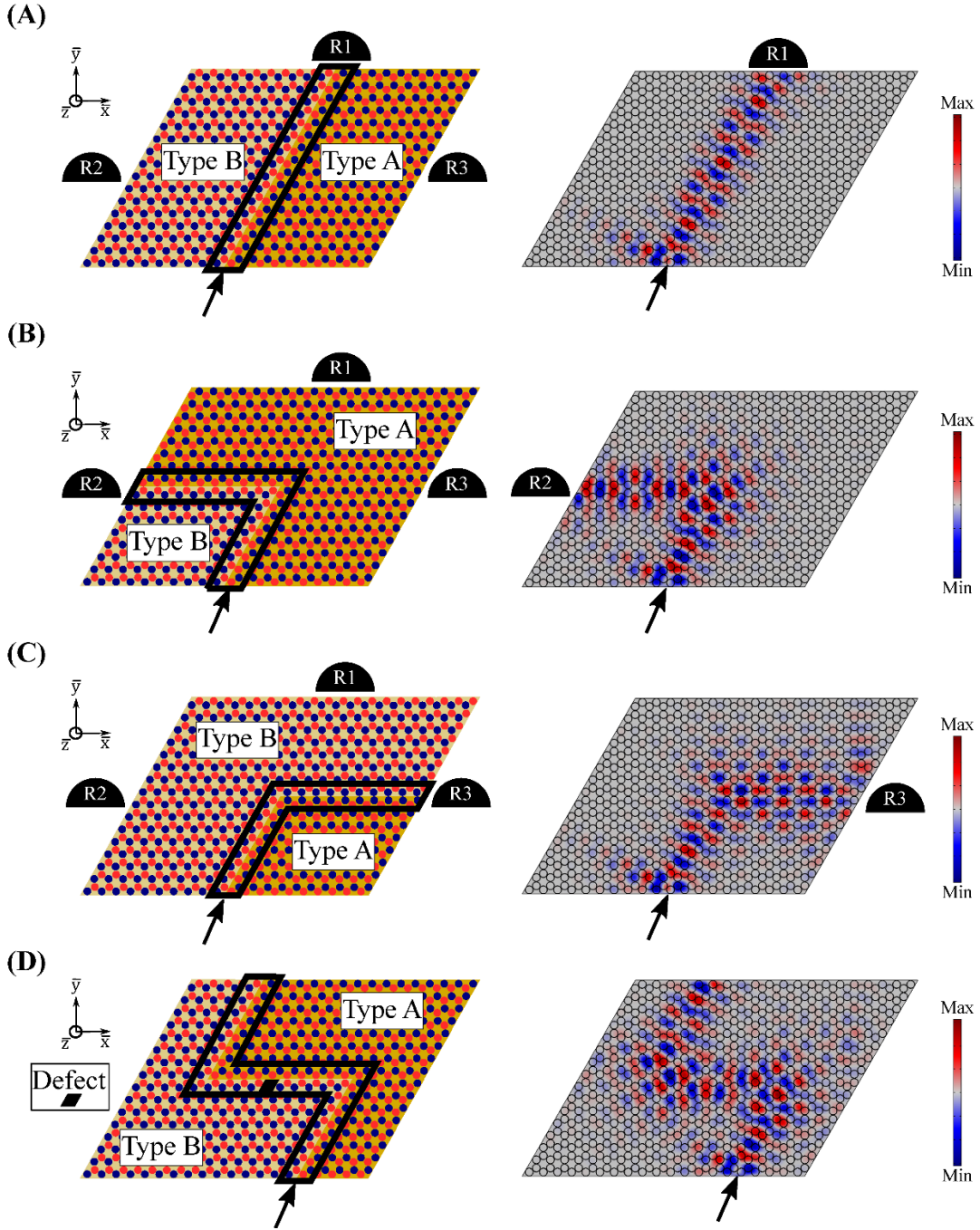

**Figure 5:** Schematics (left column) and steady-state response (right column) for guided wave propagation along **(A)** straight, **(B)** sharp corner, **(C)** shallow corner, and **(D)** 'Z-shape with defect' interfaces. In the schematics, R1, R2, and R3 represent output signal receivers and black lines enclose the interface between Type A and Type B lattices. Circuit parameters are defined as: $\beta = \pm 0.04$, $\Omega_{t-eff} = 16.0$, and $\xi = 0.79$, resulting in $\Omega_{Dirac} = 8.9$ and $\vartheta_{eff} = 2$. A harmonic out-of-plane point excitation ($\Omega_e = 8.7$) is applied where indicated by the black arrow. For steady-state response, the out-of-plane displacement amplitude is indicated by the color intensity. The steady-state displacement fields illustrate the path tunability and topological protection of the proposed metamaterial.

## 4.1 Frequency Tunability of the Dirac Point

Since topological wave propagation occurs at a frequency near the Dirac frequency ($\Omega_{Dirac}$), the tunability of the Dirac point is investigated. The eigenvalue problem derived from the PWE method (**Equation 13**) is utilized to calculate $\Omega_{Dirac}$ as a function of the effective circuit tuning frequency $\Omega_{t-eff}$. Calculations are conducted with $\beta = 0$ for three different effective electromechanical coupling values that are representative of cases where: PZT-5H is used without negative capacitance circuitry ($\xi = 0$, $\vartheta_{eff} = 0.42$), an advanced piezoceramic (PNN-PZT developed by Gao et al., 2018) with higher material coupling is utilized ($\xi = 0$, $\vartheta_{eff} = 0.56$), and negative capacitance is implemented with PZT-5H ($\xi = 0.79$, $\vartheta_{eff} = 2$). **Figure 6A** illustrates how $\Omega_{Dirac}$ can be continuously tuned through a large frequency range ($\Omega_{Dirac}$ exists between 0 and 17.55) by tailoring the effective tuning frequency $\Omega_{t-eff}$ of the resonant circuits. Alterations to the lattice constant $a$, which can be difficult or impractical to achieve in mechanical structures after they have been fabricated, are not required to tune the Dirac frequency, since the proposed metamaterial relies on locally resonant circuits instead of a Bragg scattering mechanism. Further observation of **Figure 6A** indicates that the Dirac frequency $\Omega_{Dirac}$ is always less than the effective tuning frequency $\Omega_{t-eff}$ of the resonant circuits (regardless of $\vartheta_{eff}$), which can facilitate the achievement of Dirac points in a deep subwavelength regime ($\Omega_{Dirac} \to 0$ for $\Omega_{t-eff} \to 0$) and aligns with results that have been previously reported for thin plate metamaterials with spring-mass resonators (Torrent et al., 2013; Zhang et al., 2020). In contrast to previous studies that have used mechanical resonators, the frequency of the Dirac point ($\Omega_{Dirac}$) in the proposed metamaterial can easily be adjusted on-line by using tunable circuit parameters ($\Omega_{t-eff}$). Therefore, the tunability of the proposed metamaterial could be exploited to achieve topological phenomena that are adaptive to variable frequency requirements derived from operating conditions or external stimuli.

According to **Figure 6A**, the achievable frequency range for $\Omega_{Dirac}$ begins at $\Omega_{Dirac} \approx 0$ and asymptotically approaches $\Omega_{Dirac} = 17.55$ as $\Omega_{t-eff}$ is made very large. This frequency range is unaltered by variation of the effective electromechanical coupling ($\vartheta_{eff}$), as can be seen in **Figure 6A** and Section 4 of the Supplementary Material, where a parametric analysis is conducted with extreme $\vartheta_{eff}$ values. Thus, 17.55 is designated as the limiting frequency ($\Omega_{limit}$) for the Dirac point (see dotted black line in **Figure 6A**). Insight into why this limit exists is gained by studying the evolution of the band structure with increasing $\Omega_{t-eff}$ (**Figures 6B,C,D**). In **Figure 6B**, the first and second dispersion curves are plotted for $\Omega_{t-eff}$ values that are between 2 and 150, with the arrow indicating how the bands evolve with increasing $\Omega_{t-eff}$. As can be seen in this figure, the Dirac point (the degeneracy between the first and second bands at $K$) converges to $\Omega_{limit} = 17.55$ (indicated with dashed red lines) for large $\Omega_{t-eff}$ values. **Figure 6C** contains the variation of the third dispersion curve for the same $\Omega_{t-eff}$ range. The band evolution illustrated in **Figure 6C** illuminates how the frequency value of the third band at the $K$ point remains constant at $\Omega_{limit} = 17.55$ for all $\Omega_{t-eff}$, and thus cannot be adjusted by circuit parameters. This phenomenon causes the third band to effectively act as a "constraint" that limits the maximum frequency of the Dirac point. A physical explanation for this observation is obtainable by analysis of **Figure 6D**, where the band structure for the proposed metamaterial with $\Omega_{t-eff} = 131$ (solid red lines) is compared to the band structure for the bare bimorph plate (dashed black lines). The two band structures are effectively identical because the tuning frequency has been set to such a large value that the effects of the resonant circuits are no longer present in this frequency region. Thus, for large $\Omega_{t-eff}$, the dispersion properties of the proposed metamaterial will converge to those of the bare bimorph plate within the targeted frequency range (the range containing the first and second bands), resulting in an "upper bound" for the Dirac frequency. Although it has yet to be reported in previous investigations, this phenomenon is also observable for a thin plate with attached spring-mass resonators. The discovery of the limiting frequency $\Omega_{Dirac}$ presented in this work and the awareness of the underlying phenomena that define it could be utilized as part of a design framework in future studies that leverage local (electrical or mechanical) resonators to achieve Dirac dispersions.

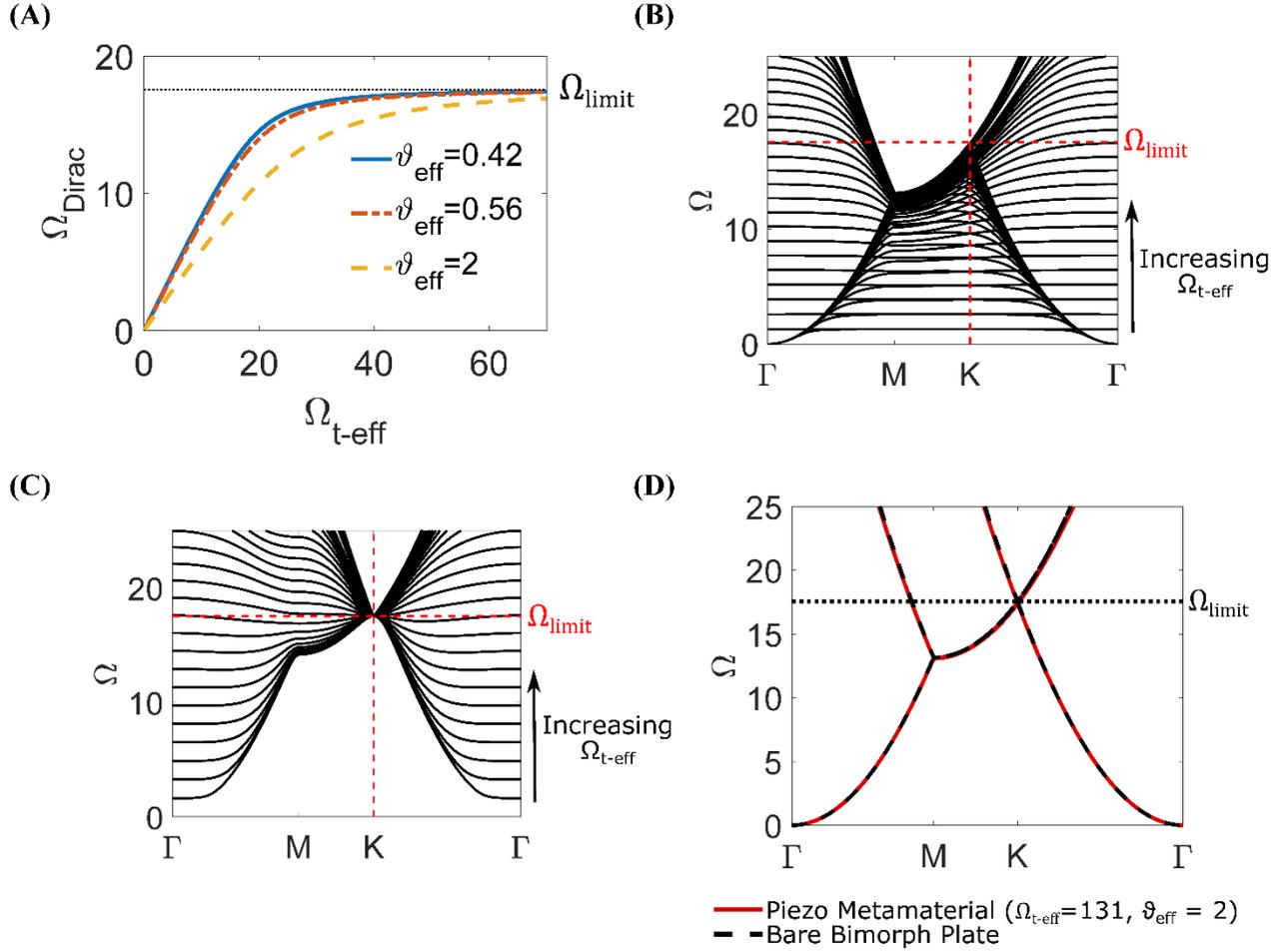

**Figure 6:** **(A)** Dirac frequency $\Omega_{Dirac}$ as a function of circuit tuning frequency $\Omega_{t-eff}$ for various $\vartheta_{eff}$. The upper limiting frequency ($\Omega_{limit}$) of 17.55 is indicated by a dotted black line. **(B)** Evolution of the first and second bands (all bands are shown as black lines) with increasing $\Omega_{t-eff}$ (indicated by arrow) for $\Omega_{t-eff}$ specified between 2 and 150 ($\beta = 0$ and $\vartheta_{eff} = 2$). $\Omega_{limit}$ is indicated by dashed red lines. **(C)** Evolution of third band for same set of parameters. **(D)** Band structure comparison for proposed metamaterial with high tuning frequency ($\Omega_{t-eff} = 131$, indicated by solid red lines) and bare bimorph plate (indicated by dashed black line).

### 4.2 Achievable Operating Region for Topological Interface States

A lattice perturbation must be applied to open a topological bandgap from the Dirac point and achieve topological edge states per the QVHE. While the previously presented analysis (see section **Working Principle – Obtainment of Tunable Topological Wave Propagation**) demonstrates this working principle under a fixed set of parameters, further exploration is required to fully evaluate the tunability of the proposed metamaterial. Thus, a parametric study involving the inductance perturbation parameter $\beta$ and the circuit tuning frequency $\Omega_{t-eff}$ is conducted to define an achievable operating region where adaptive topological wave propagation could exist. The classical eigenvalue problem for the unit cell (**Equation 13**) is solved for a wide range of $\beta$ and $\Omega_{t-eff}$ values, and the achievable operating region is defined by evaluating the previously specified performance criteria of (1) a nontrivial potential operating bandwidth, as suggested by $\Omega_{bandgap-relative} > 0$, and (2) a sufficient level of topological protection, as indicated by the valley Chern number ($|C_{v-p}| \geq 0.3$ for $p = 1,2$). For this parametric study, all geometric and material parameters are selected as indicated in **Table 1**, and no negative capacitance is connected ($\xi = 0$), such that $\vartheta_{eff} = 0.42$. The magnitude of the valley Chern number ($|C_{v-p}|$) and the relative bandgap size ($\Omega_{bandgap-relative}$) are calculated and shown in **Figures 7A,B** as a function of the lattice perturbation magnitude $|\beta|$ and the Dirac frequency $\Omega_{Dirac}$ (which is determined by $\Omega_{t-eff}$, as shown in the preceding section). The magnitude of the valley Chern number is listed more generally as $|C_v|$ in **Figure 7A**, because calculations indicate that: $C_{v-1}^{Type\,A} = C_{v-2}^{Type\,B} \approx -C_{v-2}^{Type\,A} = -C_{v-1}^{Type\,B}$ ∴ $|C_{v-p}^{Type\,A}| \approx |C_{v-p}^{Type\,B}| = |C_v|$ for $p = 1,2$. In **Figure 7A**, it is shown that the localization of the Berry curvature $|C_v|$ decreases as the inductance perturbation $|\beta|$ is increased for a fixed $\Omega_{Dirac}$. On the other hand, results in **Figure 7B** illustrate how $\Omega_{bandgap-relative}$ increases as $|\beta|$ is made larger for a particular $\Omega_{Dirac}$. This tradeoff between topological protection and potential operating bandwidth is a common feature of elastic metamaterials mimicking the QVHE, and previous works have investigated how to overcome the limitations associated with balancing these performance criteria (Du et al., 2020; Nguyen et al., 2019). Due to the adaptivity of the proposed metamaterial, the frequency range where topological wave propagation is achievable is not restricted to the potential operating bandwidth under fixed parameters. The frequency range where topological interface states that satisfy performance criteria (1) and (2) are attainable is indicated by the black dashed lines in **Figure 7B**. This enclosed region is referred to as the "achievable operating region," which is dependent on the predefined performance criteria. For this study, the achievable operating region spans $\Omega_{Dirac} = 4.2$ to $\Omega_{Dirac} = 11$, and examples of topological interface states derived from Dirac points at $\Omega_{Dirac} = 5.9, 8.9,$ and $10.4$ are shown in **Figure 7C** (corresponding $|\beta|$, $\Omega_{Dirac}$ are marked with ●, ■, ▲, ★, and ⬟ in **Figure 7B**). These topological interface states are obtained by creating a Type I interface in a finite strip of 18 unit cells (see schematic in **Figure 7C**) and selecting the most localized interface mode ($\lambda_m$, see **Equation 20**) present in the topological bandgap of the dispersion diagram. The results illuminate how the frequency tunability of the proposed metamaterial could be utilized to achieve topological interface states over a broadband set of frequencies. In addition, the evaluation of the achievable operating region reveals that while Dirac points can be obtained with a $\Omega_{Dirac}$ between 0 and 17.55, topological interface states are attainable over a narrower frequency range (for the performance criteria specified in this case: $\Omega_{Dirac} = 4.2$ to $\Omega_{Dirac} = 11$).

Aside from enabling an adjustable frequency range for topological interface states, the proposed metamaterial also facilitates the tailoring of interface mode shape and localization. The ability to adjust interface mode localization is investigated by varying the inductance perturbation parameter $|\beta|$ for a fixed $\Omega_{Dirac}$ within the achievable operating region. For each $|\beta|$, the interface mode shapes that contain maximum displacement localization ($\lambda_m$) within the topological bandgap are selected. For $\Omega_{Dirac} = 8.9$, $|\beta|$ is set to 0.005, 0.010, and 0.015, and the corresponding interface modes are shown in **Figure 7C** (marked by ■, ▲, and ★, respectively). Analysis of the mode shapes indicates that the displacement localization at the interface increases substantially as $|\beta|$ is made larger, from $\lambda_m = 0.34$ for $|\beta| = 0.005$ (■ in **Figure 7C**) to $\lambda_m = 0.69$ for $|\beta| = 0.015$ (★ in **Figure 7C**). This increased mode localization is due to a progressively enhanced band inversion between the Type A and Type B unit cells that make up the interface. In addition to tailoring the displacement localization of the interface state, the mode shape could be switched from symmetric to antisymmetric by changing to the appropriate interface type (Type I or Type II, see **Figure 4**). This capability to manipulate the interface mode could be leveraged in applications requiring adjustable displacement fields.

### 4.3 Influence of Electromechanical Coupling on Topological Wave Tunability

The effect of electromechanical coupling on the vibration attenuation performance of piezoelectric metamaterials is well documented (Hagood and von Flotow, 1991; Sugino et al., 2017; Tang and Wang, 2001). In this investigation, the influence of the effective electromechanical coupling factor $\vartheta_{eff}$ on the tunability of topological edge states in piezoelectric metamaterials is investigated for the first time. In the preceding section, a framework for the evaluation of topological edge state tunability was synthesized and applied to a baseline example case (piezoelectric layers made of PZT-5H, $\xi = 0$, $\vartheta_{eff} = 0.42$). In this section, two additional cases with identical geometric parameters are analyzed to assess the effect of enhanced electromechanical coupling: PNN-PZT piezoelectric layers with no negative capacitance circuitry ($\xi = 0$, $\vartheta_{eff} = 0.56$) and PZT-5H piezoelectric layers with a parallel-connected negative capacitor ($\xi = 0.79$, $\vartheta_{eff} = 2$). In **Figures 8A,B**, the relative bandgap $\Omega_{bandgap-relative}$ is shown as a function of the inductance perturbation parameter ($|\beta|$) and the Dirac frequency ($\Omega_{Dirac}$) for cases where $\vartheta_{eff} = 0.56$ and $\vartheta_{eff} = 2$, respectively. Dashed black lines enclose the achievable operating regions where the performance criteria ($\Omega_{bandgap-relative} > 0$ and $|C_{v-p}| \geq 0.3$) are satisfied. By comparison of **Figure 7B** with **Figures 8A,B**, it is apparent that the achievable operating region is augmented by enhancing the electromechanical coupling. This expansion occurs because the band structure is "shaped" in a manner that is beneficial from the perspective of the QVHE when the electromechanical coupling is increased (see Section 5 of the Supplementary Material for further information).

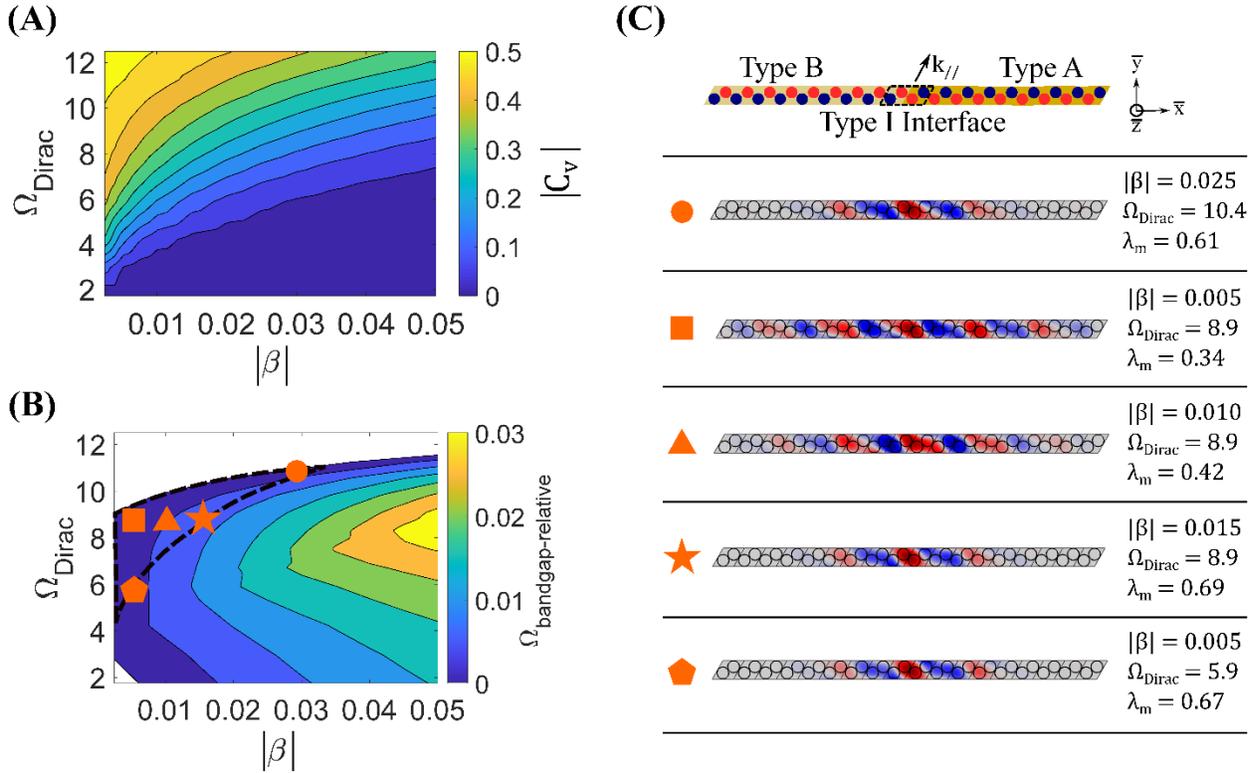

**Figure 7:** **(A)** Valley Chern magnitude $|C_v|$ calculated as a function of Dirac frequency $\Omega_{Dirac}$ and inductance perturbation magnitude $|\beta|$ with $\vartheta_{eff} = 0.42$. Increasing $|C_v|$ indicated by increasing brightness. **(B)** Relative bandgap $\Omega_{bandgap-relative}$ as a function of Dirac frequency $\Omega_{Dirac}$ and inductance perturbation magnitude $|\beta|$ with $\vartheta_{eff} = 0.42$. Increasing $\Omega_{bandgap-relative}$ indicated by increasing brightness. The achievable operating region is enclosed by the dashed black line. **(C)** Schematic for finite strip with Type I interface that is used to generate interface modes. Interface mode shapes for finite strip with unit cell parameters indicated by ●, ■, ▲, ★, and ⬟ markings in **(B)**.

The expansion of the achievable operating region signifies an increased level of tunability. In terms of frequency range, the achievable operating region is extended to lower frequencies (while maintaining an upper bound of approximately $\Omega_{Dirac} = 11$), as the lower boundary extends to $\Omega_{Dirac} = 3.8$ for $\vartheta_{eff} = 0.56$ and $\Omega_{Dirac} = 3.2$ for $\vartheta_{eff} = 2$, resulting in 6% and 15% increases in operating region frequency range, respectively (when compared to $\Omega_{Dirac} = 4.2$ to $11$ for $\vartheta_{eff} = 0.42$). In addition, for any selected $\Omega_{Dirac}$, a greater $|\beta|$ can be attained (i.e., the right edge of the achievable operating region is extended to larger $|\beta|$ values). As demonstrated in the previous section, a larger $|\beta|$ corresponds to a wider potential operating bandwidth under fixed parameters ($\Omega_{bandgap-relative}$) and an amplified band inversion at the interface. Due to this increased band inversion, the topological interface mode shapes for the $\vartheta_{eff} = 0.56$ and $\vartheta_{eff} = 2$ cases (shown in **Figures 8C,D**) contain greater localization at the interface than the interface modes for $\vartheta_{eff} = 0.42$ (**Figure 7C**). Regardless of the specified $\Omega_{Dirac}$ (examples shown in **Figures 7,8** are $\Omega_{Dirac} = 5.9, 8.9$, and $10.4$), the greatest achievable mode localization, which is quantified by $\lambda_m$, increases with the effective coupling (e.g., for $\Omega_{Dirac} = 8.9$, $\lambda_m = 0.69, 0.74, 0.92$ when $\vartheta_{eff} = 0.42, 0.56, 2$, respectively, see ★ in the **Figures 7C,8C,D**).

These findings indicate that electromechanical coupling plays a crucial role in determining the extent of the tunability of the topological interface state. Thus, care must be taken to maximize the effective electromechanical coupling through material selection and geometric design. In addition, negative capacitance circuitry could be utilized to artificially enhance the coupling and achieve topological interface states with a broader frequency range and increased displacement localization.

**4.4 Finite Element Evaluation of Frequency and Mode Shape Tunable Topological Waves**

FE simulations of plates constructed from the proposed metamaterial are conducted to verify that tunable topological wave propagation can be realized by activation of the topological interface states contained within the achievable operating region. Plates with straight line and Z-shaped Type I interfaces are harmonically excited at the locations indicated by the arrows in **Figure 9A**. Geometric, material, and negative capacitance ($\xi = 0.79$) parameters are selected to match the analysis presented in section **Working Principle – Obtainment of Tunable Topological Wave Propagation**, where wave propagation for $|\beta| = 0.04$, $\Omega_{Dirac} = 8.9$ (★ in **Figures 8B,D**), and $\Omega_e = 8.7$ was displayed (see **Figure 5**). To complement the results shown for $\Omega_{Dirac} = 8.9$, topological wave propagation is investigated for high frequency ($\Omega_{Dirac} = 10.4$) and low frequency ($\Omega_{Dirac} = 5.9$) Dirac points that are near the upper and lower boundaries of the achievable operating region defined in **Figure 8B**. **Figure 9B** contains the steady-state displacement fields for a plate with circuit parameters specified as $|\beta| = 0.050$ and $\Omega_{Dirac} = 10.4$ (● in **Figures 8B,D**) that is harmonically excited at $\Omega_e = 10.2$. The flexural displacement is successfully guided along the interfaces. However, displacement localization at the interface is noticeably reduced from what is observed in **Figure 5** due to a less localized interface mode ($\lambda_m = 0.75$ for ● and $\lambda_m = 0.92$ for ★ in **Figure 8D**). The steady-state displacements for $|\beta| = 0.032$, $\Omega_{Dirac} = 5.9$ (⬣ in **Figures 8B,D**), and $\Omega_e = 5.9$ are shown in **Figure 9C**. The flexural displacement is highly localized to the interface for both the straight line and Z-shaped cases. These high- and low-frequency results, paired with the results for $\Omega_{Dirac} = 8.9$, illustrate how guided topological wave propagation in the proposed metamaterial can be tuned across a broad frequency range. The adjustment of the displacement field (i.e., localization) at the interface is also displayed.

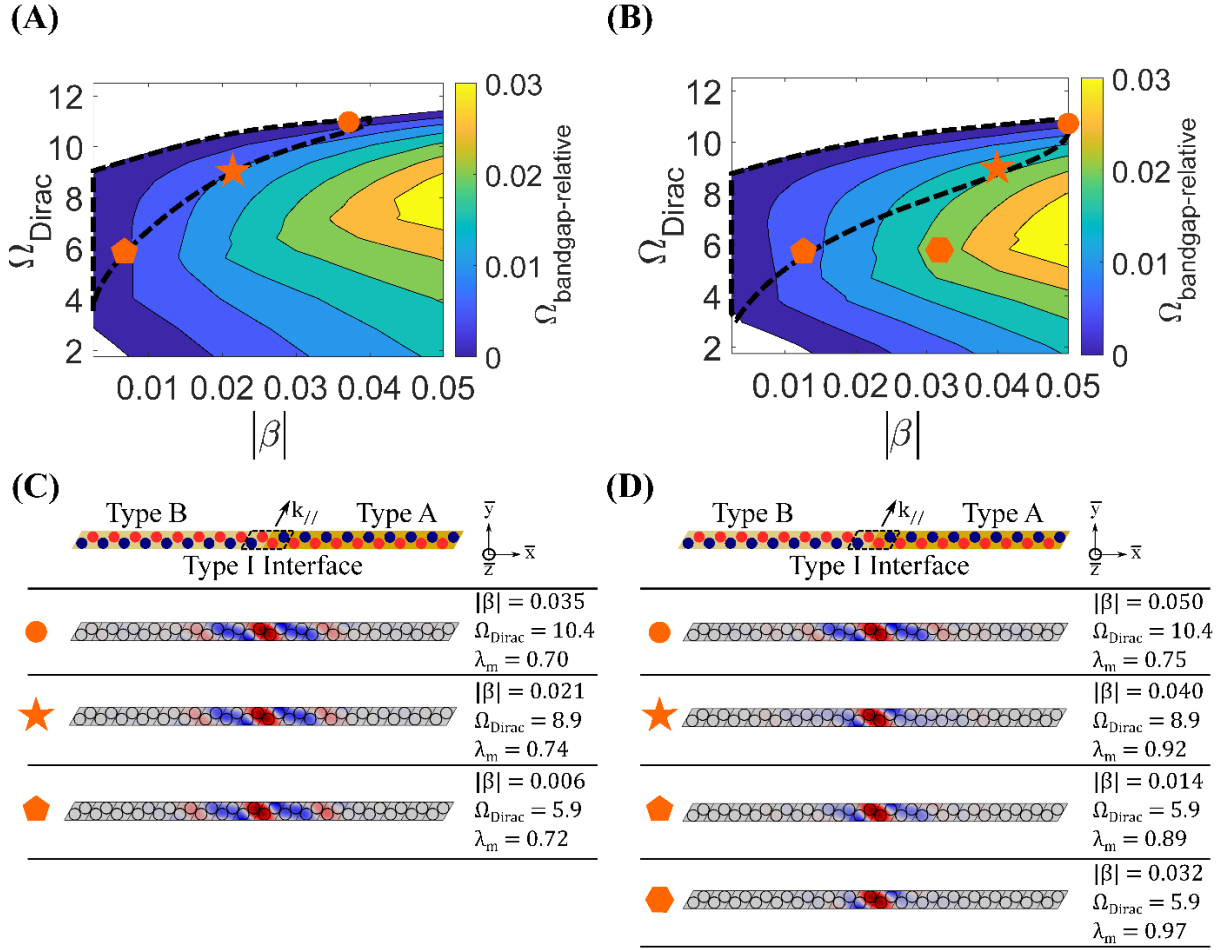

**Figure 8: (A)** Relative bandgap $\Omega_{bandgap-relative}$ as a function of Dirac frequency $\Omega_{Dirac}$ and inductance perturbation magnitude $|\beta|$ for PNN-PZT piezoelectric layers ($\vartheta_{eff} = 0.56$). Increasing $\Omega_{bandgap-relative}$ indicated by increasing brightness. The achievable operating region is enclosed by the dashed black lines. **(C)** Schematic for finite strip with Type I interface that is used to generate interface modes. Interface mode shapes for finite strip with unit cell parameters indicated by ●, ★, ⬟, and markings in **(A)**. **(B)** and **(D)** report the same results as **(A)** and **(C)**, respectively, for PZT-5H with negative capacitance circuitry ($\vartheta_{eff} = 2$). Findings illustrate how the size of the achievable operating region, and thus tunability of the topological interface state, increases with larger $\vartheta_{eff}$.

Notably, for the low-frequency case ($\Omega_{Dirac} = 5.9$), the inductance perturbation ($|\beta| = 0.032$) is outside of the achievable operating region (⬣ in **Figures 8B,D**). For the maximum $|\beta|$ contained within the achievable operating region ($|\beta| = 0.014$, ⬟ in Figure **8B,D**), the flexural response does not successfully maintain localization along the sharp corners in the lattice with the Z-shaped interface (see Section 6 of the Supplementary Material). Thus, despite the resulting reduction in valley Chern magnitude ($|C_v| = 0.015$ for $|\beta| = 0.032$), a larger $|\beta|$ (and corresponding augmented band inversion and topological bandgap) is required to successfully guide a wave along multiple sharp corners within this deep subwavelength frequency regime. This result, paired with a similar observation reported in (Nguyen et al., 2019), indicates that under certain conditions (e.g., multiple sharp corners at very low frequencies), the performance criteria that define the achievable operating region may need to be modified (e.g., reducing the minimum $|C_v|$ requirement). Due to the comprehensive tunability of the proposed metamaterial, circuit adjustments could be made on-line to achieve the desired performance (as is shown in **Figure 9C**) when scenarios such as these arise (see Section 6 of the Supplementary Material for further discussion).

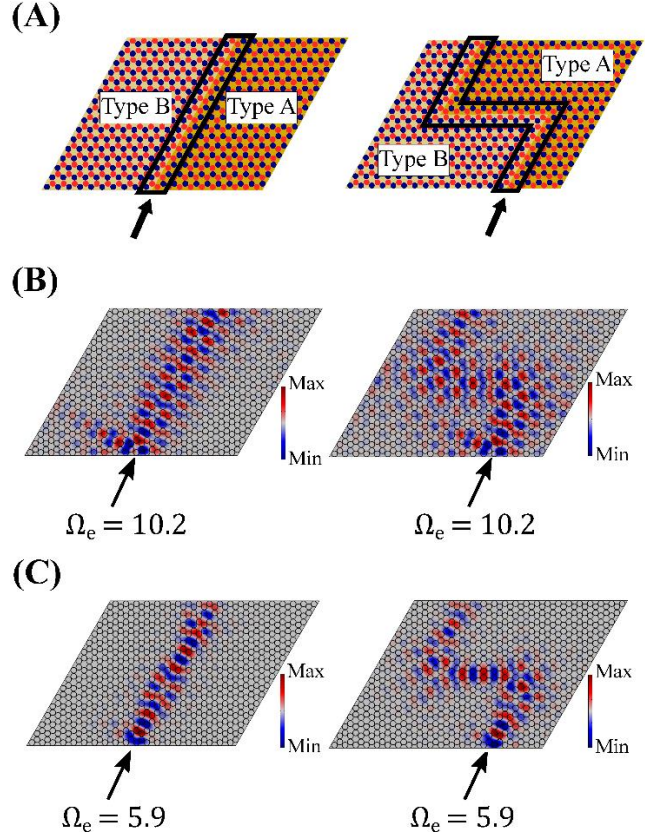

**Figure 9:** **(A)** Schematics for thin plate metastructures with straight and Z-shaped Type I lattice interfaces (enclosed in black lines). A harmonic out-of-plane point excitation is applied where indicated by the arrow. Negative capacitance circuity is connected such that $\vartheta_{eff} = 2$ in all cases. **(B)** Steady-state displacement fields for $|\beta| = 0.050$, $\Omega_{Dirac} = 10.4$, and $\Omega_e = 10.2$. **(C)** Steady-state displacement fields for $|\beta| = 0.032$, $\Omega_{Dirac} = 5.9$, and $\Omega_e = 5.9$. For steady-state response, the out-of-plane displacement amplitude is indicated by the color intensity. The results illuminate how tunable topological interface states can be harnessed to achieve topological wave propagation with adjustable levels of displacement localization over a large frequency range

## 5 Lattice Reconfiguration

The architecture of the proposed metamaterial can also be exploited to enhance adaptivity through lattice reconfiguration, which has been used in previous investigations to achieve tailorable bandgaps and waveguides in topologically trivial structures (Thota et al., 2017; Thota and Wang, 2018). In this section, lattice reconfiguration is explored as a mechanism to enhance the frequency range of topological waves and obtain additional topological edge states.

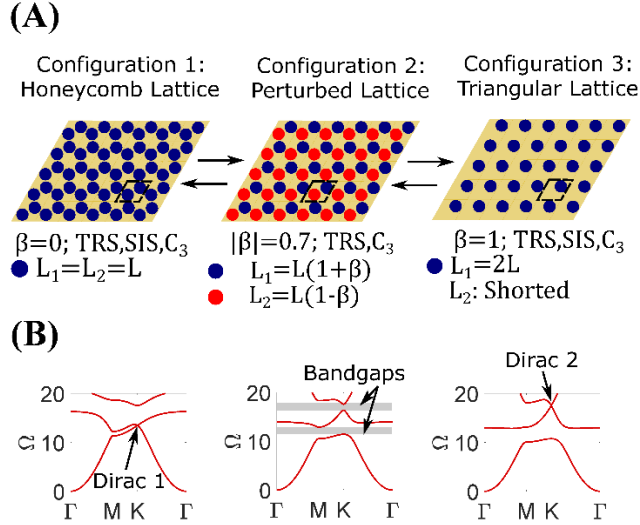

**Figure 10:** **(A)** Schematics for three lattice configurations attainable through circuit tailoring in the proposed metamaterial. Circuits are connected to electrodes that are represented by blue and red circles. For Configuration 3, electrodes pertaining to shorted circuits are omitted for clarity. Unit cell (enclosed in black dashed lines) circuit parameter details and lattice symmetries are listed below each schematic. **(B)** Dispersion diagrams corresponding to lattice Configurations 1-3 calculated using the PWE method with $\Omega_{t-eff} = 28$ and $\vartheta_{eff} = 2$. Dirac 1, topological bandgaps, and Dirac 2 are all highlighted for Configuration 1, Configuration 2, and Configuration 3, respectively.

## 5.1 Formation of Additional Dirac Point Through Lattice Reconfiguration

**Figure 10A** contains schematics for three different lattice configurations that are attainable with the proposed metamaterial. **Figure 10B** contains the corresponding dispersion diagrams for each lattice configuration with circuit conditions specified such that $\Omega_{t-eff} = 28$ and $\vartheta_{eff} = 2$. Configuration 1 is a honeycomb lattice that contains identical resonant circuits ($\beta = 0$) and the symmetries ($C_3$, SIS, TRS) required to achieve a Dirac point between the first and second bands (labeled as "Dirac 1" in **Figure 10B**). Configuration 3 is a triangular lattice that is realized by shorting one of the two circuits in the unit cell ($\beta = 1$, see **Figure 10A**). This triangular lattice also contains $C_3$, SIS, and TRS symmetries, resulting in the formation of a Dirac point between the second and third bands at $\Omega_{Dirac}$ = 17.55 (labeled as "Dirac 2" in **Figure 10B**). Previous works concerning triangular lattices in photonic crystals have also uncovered a Dirac point between the second and third bands (Kim et al., 2013; Plihal and Maradudin, 1991; Zhang, 2008), supporting this result. As shown in previous sections, Dirac 1 can be tuned from $\Omega_{Dirac} = 0$ to $\Omega_{Dirac} = 17.55$ due to the resonant characteristic of the connected circuitry. In addition, topological interface states derived from Dirac 1 are attainable for a narrower frequency range that covers $\Omega_{Dirac} = 3.2$ to $\Omega_{Dirac} = 11$. Unlike Dirac 1, the formation of Dirac 2 is the result of a Bragg scattering mechanism, and it is not frequency tunable with resonant circuit parameters. However, it exists at a high frequency ($\Omega_{Dirac} = 17.55$) that is outside of the operating range for topological interface states derived from Dirac 1, and thus could be exploited to further broaden the frequency range of interface states in the proposed metamaterial. Furthermore, this additional Dirac point could be employed to achieve other capabilities derived from the rich physics associated with Dirac cones, such as boundary states.

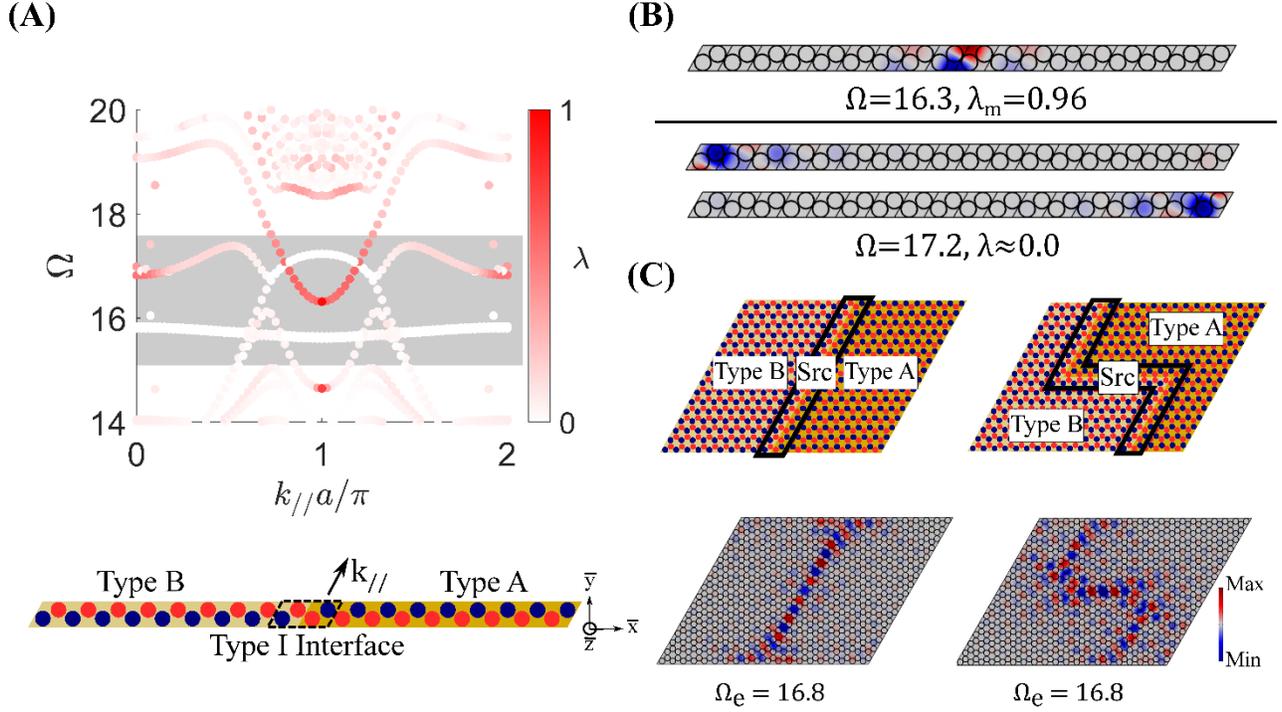

**Figure 11: (A)** Band structure for a finite strip ($|\beta| = 0.70$, $\Omega_{t-eff} = 28.0$, $\vartheta_{eff} = 2$) with a Type I interface. Dark red shading indicates localized interface states ($\lambda \approx 1$) and white shading indicates boundary states ($\lambda \approx 0$). The rectangular gray shaded region represents a frequency range where no bulk modes exist. **(B)** A symmetric mode shape that is calculated from the interface state at $\Omega = 16.3$ with a localized displacement ($\lambda_m = 0.96$) is shown. Two boundary mode shapes that are calculated from the degenerate boundary states at $\Omega = 16.8$ ($\lambda = 0$) are also shown. **(C)** (top row) Schematics for thin plate metastructures with straight and Z-shaped Type I lattice interfaces (enclosed in black lines). A harmonic ($\Omega_e = 16.8$) out-of-plane point excitation is applied where indicated by 'Src'. (bottom row) Steady-state displacement fields illustrating guided wave propagation for the high-frequency interface states.

## 5.2 High-Frequency Interface State from Dirac 2

The same process described in previous sections for Dirac 1 is followed to construct interface states from Dirac 2. **Figure 10A** includes the schematic for a lattice (defined as Configuration 2: Perturbed Lattice) with the inductance perturbation parameter $\beta$ specified to be between 0 and 1 ($|\beta| = 0.7$), which breaks SIS. **Figure 10B** contains the dispersion diagram for Configuration 2, where it is shown that bandgaps exist in place of Dirac 1 and Dirac 2. For Dirac 2, the bandgap extends from a fixed upper boundary of $\Omega = 17.55$ to lower frequencies ($\Omega_{bandgap}$ spans $\Omega = 15.13$ to $\Omega = 17.55$ for $|\beta| = 0.7$). To investigate whether an interface state can be obtained within this bandgap, Type A ($\beta = 0.7$) and Type B ($\beta = -0.7$) lattices are connected at a Type I interface, and a dispersion analysis is conducted for a finite strip of 18 unit cells (see schematic in **Figure 11A**). For this analysis, periodic boundary conditions are applied in the $k_{//}$ direction, and the ends are specified as free. The dispersion diagram for the finite strip is shown in **Figure 11A**, where it is apparent that a localized interface state exists within the bandgap (the interface state is the red band present in the gray shaded bandgap). A symmetric and highly localized ($\lambda_m = 0.96$) interface mode shape evaluated at $\Omega = 16.3$ is shown in **Figure 11B**. In addition to the interface state, multiple localized edge states exist at the left and right boundaries of the finite strip, as indicated by the white circles in the band diagram (for $\lambda \approx 0$, dispersion curves are white, **Figure 11A**). Mode shapes for two degenerate "boundary states" with flexural displacement localized at the left and right boundaries are shown for $\Omega = 17.2$ in **Figure 11B**. Similar results for a Type II interface and the corresponding antisymmetric interface state are contained within Section 7 of the Supplementary Material.

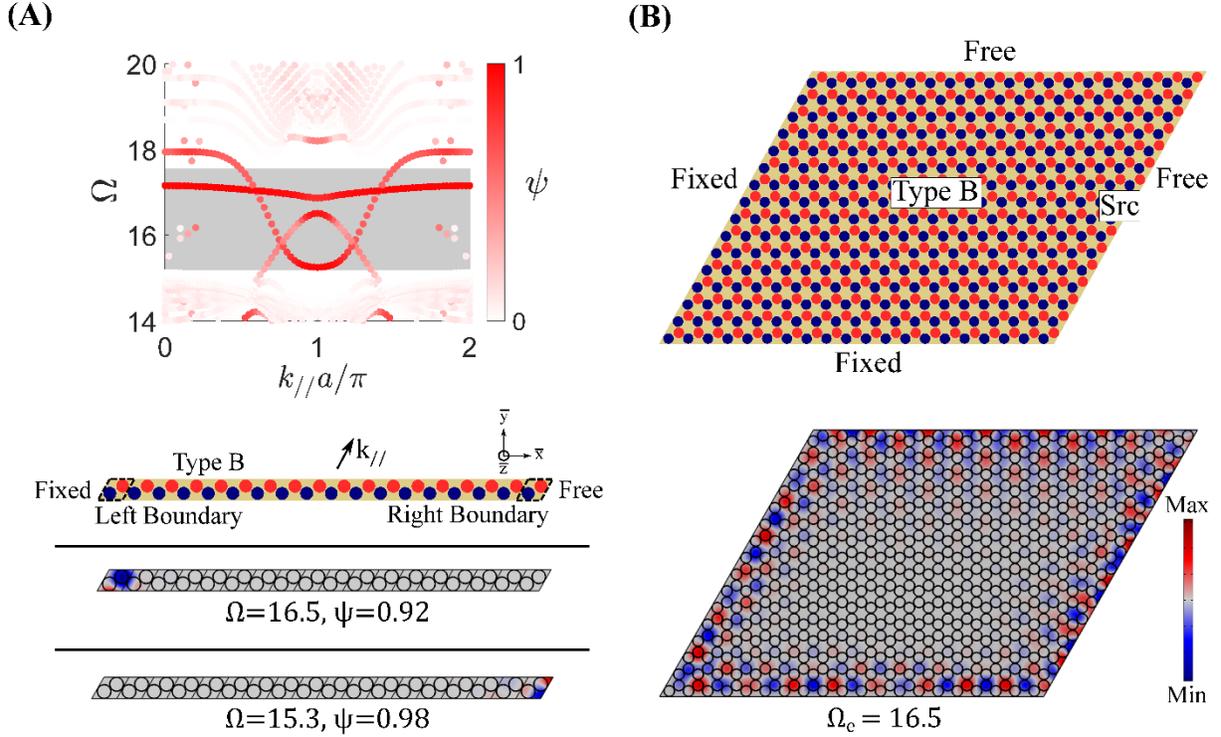

**Figure 12: (A)** Band structure for a finite strip ($|\beta| = 0.70$, $\Omega_{t-eff} = 28.0$, $\vartheta_{eff} = 2$) comprised of Type B unit cells (no interface, see schematic). Dark red shading indicates localized boundary states ($\psi \approx 1$). The rectangular gray shaded region indicates a frequency range where no bulk modes exist. Left and right boundary mode shapes are shown. **(B)** (top) Schematic for thin plate metastructure with fixed and free boundaries selected to attain boundary states. A harmonic ($\Omega_e = 16.5$) out-of-plane point excitation is applied where indicated by 'Src'. (bottom) Steady-state displacement field illustrating flexural displacement confinement along plate boundaries.

To demonstrate guided wave propagation, plates are constructed with straight and Z-shaped Type I interfaces and excited harmonically ($\Omega_e = 16.8$) with a point source at the locations indicated by "Src" in **Figure 11C** (see schematics in top row of **Figure 11C**). The resulting steady-state displacement fields show that flexural displacement can be localized along the interfaces for these high-frequency interface states (see steady-state displacement fields in bottom row of **Figure 11C**). However, coupling between the interface states and boundary states does occur. Therefore, to selectively activate an interface state, care must be taken to specify an excitation (e.g., location at the center of the plate) that will not localize displacement at the boundaries.

### 5.3 Boundary States from Dirac 2

Results from the finite strip dispersion analysis for a Type I interface (**Figure 11**) suggest that boundary states are readily achievable from Dirac 2. To further investigate the formation of boundary states from Dirac 2, a dispersion analysis is conducted for a finite strip with 18 Type B unit cells (no interface is present, see schematic in **Figure 12A**). The left boundary is specified as fixed, the right boundary is specified as free, and periodic boundary conditions are applied in the $k_{//}$ direction. The left boundary is fixed because a fixed condition enforces dynamic behavior at the boundary that closely approximates the interface behavior for the antisymmetric interface state. As shown in Section 7 of the Supplementary Material, the antisymmetric interface state is obtainable by creating a Type II interface. Since the Type B unit cell at the left boundary comprises the right half of a Type II interface, the fixed condition effectively approximates the left half, and a boundary state that resembles half of the antisymmetric interface state is expected to occur (Liu and Semperlotti, 2018). Similarly, a boundary state closely approximating half of the symmetric interface state is expected to appear at the right boundary, which is designated as free. For each band in the calculated dispersion diagram, a boundary localization parameter $\psi$ is defined to measure the amount of flexural displacement that is localized at the left and right boundaries as:

$$\psi = \frac{\iiint_{V_{boundaries}} |w|^2 \, dV}{\iiint_{V_S} |w|^2 \, dV} \tag{21}$$

where $V_{boundaries}$ is the total volume of the two unit cells at the boundaries (enclosed in the dashed black boxes in **Figure 12A**) and $V_S$ is the volume of the entire finite strip. This boundary localization parameter is evaluated for each band and is represented in the dispersion diagram by color intensity. Boundary states are indicated in the dispersion diagram (**Figure 12A**) as dark red bands ($\psi \approx 1$). One boundary state emerges from a lower frequency set of bulk modes and crosses into the bandgap (gray shaded region) towards a higher frequency set of bulk modes. The mode shape for this band with the maximum level of displacement localization at the boundary is shown in **Figure 12A** ($\psi_m = 0.92$ for $\Omega = 16.5$), indicating that this band corresponds to a left boundary state that approximates half of the antisymmetric interface state derived from Dirac 2. Another boundary state emerges from the higher frequency set of bulk modes and crosses towards the lower frequency set of bulk modes. This boundary state is localized to the right boundary, as illustrated by the mode shape calculated for $\Omega = 15.3$ ($\psi_m = 0.98$), and approximates half of the symmetric interface state.

A plate is constructed with boundary conditions (free and fixed) specified such that the outlined boundary states are supported at all four boundaries. The plate is excited harmonically ($\Omega_e = 16.5$) with a point source located as indicated by "Src" in **Figure 12B**. Results show that both boundary states are activated and compatible with each other, resulting in flexural wave propagation around multiple sharp and shallow corners that is localized to all four boundaries of the plate. This capability to achieve adaptive wave propagation along structural boundaries could be beneficial in applications requiring confinement of energy along the edges of structures, such as vibration isolators.

## 6    Conclusions

This research proposes and develops a topological metamaterial that harnesses resonant piezoelectric circuitry to enable comprehensively tunable elastic wave control. Overall, this investigation advances the state of the art by enabling and exploring the adaptation of topological wave path, frequency, and edge mode shape in a single mechanical platform for the first time. The proposed metamaterial operates over a broad frequency bandwidth and can be integrated in a compact fashion for applications that require control of large-wavelength (i.e., low-frequency) waves due to its subwavelength characteristic, which constitutes a breakthrough in the field of tunable topological elastic waves.

In this manuscript, the tunability of wave path, frequency range, and edge states is explored through a systematic analysis of the dispersion properties and dynamic response characteristics of the proposed metamaterial. A subwavelength and frequency tunable Dirac point (Dirac 1) is uncovered, and it is revealed that symmetric and antisymmetric topological interface states can be obtained from it. FE simulations illuminate how these interface states can be activated to achieve guided elastic wave transmission that is robust to disorder and defects and can be manipulated on-demand into a myriad of desired directions. A deeper understanding of the adaptive characteristics of the metamaterial is formed through parametric studies. A finite tunable frequency range and its underlying physical basis are discovered for Dirac 1. An achievable operating region is defined for topological interface states derived from Dirac 1, where it is learned that interface states that meet specified performance metrics are achievable over a wide frequency bandwidth that comprises a subset of the Dirac 1 frequencies. These findings offer new insights into frequency tunability that may be leveraged in future studies concerning Dirac dispersions and the QVHE in locally resonant elastic metamaterials. Further exploration of the achievable operating region illuminates how circuit parameters can be utilized to tune the displacement field of a waveguide by tailoring the localization and shape of the interface state. The operating region is used as a framework to study the role of electromechanical coupling in topological wave propagation for the first time. Results indicate that increased electromechanical coupling enhances the frequency range and achievable interface mode localization of the interface states. Lattice reconfiguration is also investigated as a method to tailor the topological properties of the proposed metamaterial. Analysis of a triangular lattice obtained through shorting circuits reveals that a second Dirac point (Dirac 2) can be achieved in a high-frequency range that extends beyond the operating region for Dirac 1 interface states. Boundary states and additional interface states are shown to be obtainable from Dirac 2, and numerical simulations illustrate how these states can be exploited to achieve exceptional guided wave phenomena.

The outcomes from this investigation provide fundamental insights into the influence of locally resonant elements, electromechanical coupling, and lattice reconfiguration in adaptive topological metamaterials exhibiting the QVHE, presenting a basis for further exploration. The proposed topological metamaterial may be employed to achieve subwavelength elastic wave control that is robust to practical considerations (e.g., sharp corners or lattice imperfections) and adaptive in real-time to shifting operating requirements and external conditions. These beneficial features could be harnessed to improve performance and expand functionalities in a range of applications requiring adaptive and robust elastic wave control, such as vibration isolators, wave filters/multiplexers, and energy harvesters.

## 7     Conflict of Interest

The authors declare that the research was conducted in the absence of any commercial or financial relationships that could be construed as a potential conflict of interest.

## 8     Author Contributions

P.D. and K.W.W. formulated the presented idea. P.D. developed the mathematical framework, performed analytical and numerical computations, and drafted the manuscript. K.W.W. discussed results with P.D., provided research direction/advisement, and edited the manuscript.

## 9     Funding

This research is funded by the National Science Foundation under Award No. 1661568. P.D. also acknowledges financial support from the Rackham Merit Fellowship at the University of Michigan.

## 10  Acknowledgments

The authors thank Xiang Liu and Megan Hathcock at the University of Michigan Structural Dynamics and Controls Lab for their useful suggestions and participation in discussions regarding topological metamaterials and FE modeling using COMSOL Multiphysics.